\documentclass[review]{elsarticle}

\usepackage{lineno,hyperref}
\usepackage{blindtext}
\usepackage{amsmath,amsfonts}
\usepackage{booktabs}
\usepackage{float}
\usepackage{multirow}
\usepackage{subcaption}
\biboptions{super}

\journal{International Journal for Numerical Methods in Fluids}

\usepackage{numcompress}

\ifpdf
    \graphicspath{{Figs/Raster/}{Figs/PDF/}{Figs/}}
\else
    \graphicspath{{paper/Figs/Vector/}{paper/Figs/}}
\fi
\usepackage{atbegshi}
\AtBeginDocument{\AtBeginShipoutNext{\AtBeginShipoutDiscard}}

\let\oldeq\equation{}\def\equation{\par\vspace{-\parskip}\oldeq}

\begin{document}
\begin{frontmatter}

\title{A finite volume coupled level set and volume of fluid method with a mass conservation step for simulating two-phase flows}
\author[mymainaddressA]{Konstantinos G. Lyras\corref{mycorrespondingauthor}},
\address[mymainaddressA]{School of Biomedical Engineering $\&$ Imaging Sciences, King's College London, London SE1 7EU, United Kingdom}\ead{konstantinos.lyras@kcl.ac.uk}\cortext[mycorrespondingauthor]{Corresponding author}
{Corresponding author}
\author[mymainaddressA]{Jack Lee \corref{mycorrespondingauthor}}\ead{jack.lee@kcl.ac.uk}

\begin{abstract}
This paper presents a finite volume method for simulating two-phase flows using a level set approach coupled with volume of fluid method capable of simulating sharp fluid interfaces. The efficiency of the method is a result of the fact that the interface is calculated in order to satisfy mass conservation with no explicit interface reconstruction step and the mass fluxes across cell-faces are corrected to respect the recovered volume fraction. The mass-conservation correction step proposed here, is utilised using an iterative algorithm which solves a reaction-diffusion equation for the mass correction of the level set. The re-sharpened volume fraction is used for the new volumetric fluxes at each cell which are calculated through the proposed algorithm that guarantees that they satisfy mass conservation. The algorithm is not limited in representing the interface with the 0.5-contour and is applicable for arbitrary polyhedral cells. Good accuracy and mass conservation are achieved when compared to other conservative approaches. 

\end{abstract}
\begin{keyword}
level set, volume of fluid, mass conservation, two-phase flow, finite volume method
\end{keyword}
\end{frontmatter}

\section{Introduction}
\subsection{Scope}
Three-dimensional multiphase flows occur in multiple problems in biomedical, automotive, chemical and nuclear industries among others. 
In particular, the accurate numerical simulation of two-phase flows of two immiscible fluids faces the problem of resolving moving interfaces that deform under the influence of various forces such as surface tension and gravitational forces.  
Two of the most popular implicit methods for capturing the interface in interfacial flows that involve large topological changes are the volume of fluid (VOF) method ~\citep{Hirt1981,Scardovelli1999,Gueyffier1999,Scardovelli2000,Tryggvason2011,
Agbaglah2011,Ling2015,Bna2016,elias2007stabilized,greaves2006simulation,xie2020control} and the level set (LS) method \citep{Osher1988,Merriman1994,Chang1996,Sethian1996,Sethian1999,Osher2001,Enright2002,Sethian2003, Osher2006}. 
One of the most attractive features of both methods is their ability to naturally adapt to any topological changes and the calculation of curvature can be implemented through simple formulations.  
LS methods offer a powerful tool for capturing the interface for various flow regimes including dispersed and bubbly flows. LS is particularly efficient in describing interfacial dynamics involving interfaces breaking up (or colliding) automatically.
In contrast to VOF, mass loss in LS can be relatively high. This might be proven catastrophic for especially for long simulations where the error in mass conservation accumulates. 

Various studies have proposed different methodologies for improving mass conservation of the LS method \citep{Sussman2000,Olsson2005,Olsson2007,Enright2002,Raessi2012,Moghadam2016,Kinzel2018,Bahbah2019,
raees2016mass,antepara2020tetrahedral,quezada2020unstructured,qian2019improved,sato2012conservative}.
The particle level set (PLS) in \citep{Enright2002} is a semi-lagrangian level set method based on the advection of particles at both sides of the interface.
Significant improvements were reported for mass conservation in the conservative level set method in \citep{Olsson2005,Olsson2007}. The authors, in addition to the advection of the level set function $\psi$, implemented a re-initialisation step for the cells near the interface using the normal vector at the interface, $\mathbf{n}$ using a compressive flux $\psi (1-\psi) \mathbf{n}$. This correction step is applied for some iterations in steady-state $\psi$ for re-sharpening the interface. The results in both two and three-dimensional tests showed improvements in the accuracy of interface capturing. 
The coupled-level-set-volume-of-fluid (CLSVOF) method of \citep{Sussman2000} was developed considering the advection of the volume fraction and coupling it with the advection of the level set function \citep{Bourlioux1995}. This is a coupled VOF-LS approach in which the interface normals, calculated from the level set, are employed for the interface reconstruction in the VOF advection.
Based on this approach, many other variations have been proposed of coupling both VOF and LS making use of the advantages of both methods. The proposed CLSVOF methods rely on either geometrical VOF methods such as piecewise-linear interface construction/calculation (PLIC) \citep{Ling2015,Ansari2016,Balcazar2016} or algebraic methods \citep{Albadawi2013}. 
Two of the most commonly met challenges for coupling VOF and LS are the interface reconstruction in VOF and the mapping algorithm to the level set field so that it conserves mass in an efficient setting suitable for different types of meshes. Both tasks are required to accurately advect the volume fraction and level set respectively.  

\subsection{Objectives}
In this paper a conservative level set method coupled with VOF is presented for simulating three-dimensional interfaces of two immiscible fluids in the framework of finite volume method. 
The LS field is advected first solving the level set equation and then is corrected in two steps: a mass-conservation step and an algorithm for correcting the mass fluxes. 

Our approach solves a reaction-diffusion equation for finding the corrected $\psi$ that satisfies mass conservation using the obtained volume fraction from VOF. In this way the interface can be located at different levels and is not limited to only the 0.5 iso-surface. 
A novel algorithm for calculating the mass fluxes based on this level set is described. The new fluxes across all the mesh faces are required for the convection term for the velocity field and solution of the set of the discretised momentum equations. The proposed implementation provides an accurate calculation for the new fluxes and can be used for meshes with arbitrary number of faces and different finite volume methods regardless the re-initialisation correction they use.
This novel algorithm gives a level set function that leads to a smoothed volume fraction which satisfies mass conservation. Based on this volume fraction, the volumetric fluxes at each face and all the mixture properties are calculated. 
Then, the new level set is used for accurately calculating the interface curvature and the surface tension force in the momentum equation. 

One difference of this approach with other coupled level set-VOF methods is that the interface is calculated in order to satisfy the mass conservation with no explicit interface reconstruction step and then corrects the mass fluxes across cell-faces accordingly.
Although \citep{Kees2011} have shown that such an interface advection might be efficient in finite element methods, this approach has garnered no attention from other level set methods implemented in the context of finite volume methods and has not prompted any further testing especially in open-source finite volume CFD codes since its publication. Such an approach would require a proper strategy for calculating fluxes to advect the updated velocity field through the cell faces.

Here, we address this issue and choose to implement this approach in the context of finite volume method.
Our approach also demonstrates at least two times better accuracy compared to the ILSVOF method which has previously reported lower error than the isoAdvector in \citep{Lyras2020}. 
We also show that presented method can improve the standard OpenFOAM solver interFoam almost by an order of magnitude for the tests here.
Most interestingly, we show here that with this mass-conservation correction step for advecting level-set using a simple Runge-Kutta re-initialisation with the idea of \citep{Kees2011}, and the flux update algorithm, one might generally be able to avoid using dedicated re-initialisation schemes for re-sharpening level set as in \citep{Lyras2020,Hartmann2010}. The latter are generally difficult to implement and might result to slower calculations if proper cell-search algorithms are not used for re-distancing. Here, a simple second-order scheme is used for the re-distancing of level set which, when deploying the proposed algorithm leads to better accuracy.  
The solver is developed for various flow regimes that involve liquid/gas or liquid/liquid interface, and is implemented in the framework of  the open source CFD code OpenFOAM \citep{Weller1998}. 

Our work, shows an accuracy which is close or better than other PLIC-VOF methods and coupled level-set-VOF methods, including in our previous work in ILSVOF \citep{Lyras2020} which we extend with better accuracy as shown in the results. We improve the accuracy of the ILSVOF by using a fundamentally different algorithm of the level set advection: In \citep{Lyras2020} we simply do not solve any equation for level-set and use a specific second-order re-initialisation scheme to compensate for interface smearing.

The methodology is tested with different test for fluid/fluid interfaces undergoing under strong deformation, for cases where interface shape is complex including corners and for flows that gravitational force is significant. Different meshes are used for assessing the capability of the algorithm to capture the interface. The results for both mass conservation and accuracy showed that the method can accurately simulate three-dimensional two-phase flows with complicated topologies.   
   
\section{Motivation and methodology}
\subsection{Marker function for level set}
The level set function is used here to describe the interface dynamics of two immiscible fluids, fluid 1 and fluid 2. In standard level set methods, the level set function $\psi$ is defined to be a signed distance function calculated from the distance at the interface, $\Gamma$ that separates the two fluids 
\begin{equation}
|\psi| = min(\vec{x}-\vec{x_{\Gamma}})
\label{psiDefinition}
\end{equation} 
A given point $\vec{x}$ in the domain may have either positive or negative distance which defines whether it belongs in fluid 1 or fluid 2. The interface that separates the two fluids is then the set of points that have a zero-distance (zero-level). 
A step (Heaviside) function $H(\psi,t)$ is used to calculate the mixture properties that is equal to 1 if $\psi >0$ and equal to 0 if $\psi <0$. The total mass at a given time step is equal to $\int_{V}^{}\rho_i H(\psi)dV$ where $\rho_i$ is the density of the fluid $i$. 
  
The mass conservation requires the following to be mass conservative for incompressible flows 
\begin{equation}
\frac{\partial H(\psi,t)}{\partial t}+ \mathbf{u} \cdot \nabla{ H(\psi,t)} =0
\label{rhoHEqn}
\end{equation}
Then $H(\psi,t)$ resembles a smoothed volume fraction $\alpha$ of one fluid and 1-$H(\psi,t)$ for the second fluid. 
The smoothed volume fraction is used for advecting the face fluxes at each face and calculating the curvature. Regardless the expression of the Heaviside function, the transition at the interface ($\psi=0$) needs to be smooth within a layer of some cells surrounding the interface. If not advected properly, the solution might not converge with increasing mesh resolution and simulations might suffer from excessive mass loss. The total mass of the fluid depends on the solution of the appropriate value of $H$: while time $t$ increases, the mass is no-longer equal to the mass at time t=0 e.g.
\begin{equation}
\int_{0}^{t}\int_{V}^{}\rho_i H(\psi)dVdt = \int_{V}^{}\rho_i H(\psi,t)dVdt-\int_{V}^{}\rho_i H(\psi,0)dV \neq 0
\label{massLossEqn}
\end{equation}
The method presented here aims to finding the appropriate value of $\psi$ for which the interfacial values of $H(\psi,t)$ approximates the position of the interface exactly and satisfies the conservation property. This new corrected value will be denoted with $
\tilde{H}(\psi,t)$ e.g. we want to find an appropriate value $\psi$ such that each timestep ($\delta t$) we have
\begin{equation}
\int_{V}^{}\rho_i \tilde{H}(\psi,t + \delta t)dVdt = \int_{V}^{}\rho_i H(\psi,0)dV 
\label{psiConservation}
\end{equation}
Here, choosing a proper $\psi$ that is corrected with a method that guarantees the correct calculation of the face fluxes at each cell and the surface tension force in the momentum equation. A constraint-step is performed for satisfying volume fraction advection at each iteration while preserving its sharp interface with a second correction step.  

\subsection{Advection and correction of level set function}
The algorithm starts by tracking the volume occupied by one fluid (fluid 1) inside a computational cell $P$ which contains fluids 1 and 2. In two-phase flows the void fraction $\chi$ is a function of the local density and the individual densities, $\rho_1 , \rho_2$ at the local pressure, $\chi = (\rho - \rho_1)/(\rho_1 -\rho_2)$. This corresponds to the volume percentage inside each cell. In an arbitrary shaped cell that contains a dispersed flow e.g. bubbles, particles or droplets in a carrier or continuous phase, the volume fraction is defined as ~\citep{Crowe2005}
\begin{equation}
\alpha= \lim_{\delta V \rightarrow V_{P}} \frac{\delta V_{\alpha}}{\delta V} = \frac{1}{V_{P}} \int_{P}^{} \chi(x) \text{d}V 
\end{equation}
where $V_{P,} \delta V_{\alpha}$ are the cell volume and the volume of the continuous phase inside the cell in case of dispesed flows. Substituting the expression for volume fraction to the continuity equation yields an equation for the time evolution of $\alpha$. The advection of the volume fraction reads 
\begin{equation}
\alpha^{n+1}= \alpha^{n} - \frac{1}{V_{P}} \sum_{f \in N_f}^{}\mathbf{n_f} \int_{n}^{n+1}\int_{f}^{} \chi(x,t) \mathbf{u(x,t)} \text{d} \mathbf{S} d t
\label{alpha_to_be_advected}
\end{equation} 
where $\mathbf{n_f}$ is a unit vector defined for each on of the $N_f$ faces $f$ for cell $P$, such that each inner product with the area vector $\mathbf{S}$ points outwards the cell. Eq.~\eqref{alpha_to_be_advected} is solved here here for obtaining the volume fraction.    

The level set function according to Eq.\ref{psiDefinition} is a distance function that is defined wherever an interface exists. The distance function can be advected using 
\begin{equation}
\frac{\partial \psi}{\partial t}+ \mathbf{u} \cdot \nabla{\psi} =0
\label{psiEqn}
\end{equation}
where $\mathbf{u}$ is the velocity field. The convective term in the level set equation is integrated over a control volume $V_{P}$, and using Gauss theorem for converting the volume integral to a surface integral Eq.~\eqref{psiEqn} becomes
\begin{equation} 
\int_{V}^{} \nabla \cdot (\psi \mathbf{u}) \text{d}V = \oint_V \psi (\mathbf{u} \cdot \mathbf{n}) \,ds \approx \sum_{f}^{} \psi_{f}F_{f}
\end{equation}
where $\psi_{f}$ is the interpolated $\psi$ at the face $f$ of cell $P$ and $F_{f} = \mathbf{S_{f}} \cdot \mathbf{u_{f}}$ is the face flux, with $\mathbf{S_{f}}$ being the face area vector normal to the face. 

For this study we use a high-order interpolation scheme for evaluating the fluxes at cell faces from the cell-averaged $\psi$. We implement a third-order weighted essentially non-oscillatory (WENO) scheme which is used for the advection of the level set in arbitrary polyhedral cells. The main idea is to use a polynomial representation for reaching high-order accuracy for interpolating from the cell centres of each volume.
Writing the level set equation as a non-linear hyperbolic conservation law we calculate the fluxes solving the Riemann problem\cite{tsoutsanis2019stencil}. Integrating over the volume $V$ with surface $S$ Eq.\eqref{psiEqn} becomes
\begin{equation}
\frac{\text{d}}{\text{d}t}{<\psi>}+\frac{1}{V}\int_{S}^{} \mathbf{F}(\mathbf{u},\psi) \cdot \mathbf{n} \partial V = 0
\label{hyperbolicPsiGauss}
\end{equation} 
where $<\psi>$ is the cell-averaged $\psi$ in steady-state and $\mathbf{n}$ is the normal vector of the surface $\partial V$ surrounding $V$. The cell-averaged value at $x$ is calculated as
\begin{equation}
<\psi(x,t)> = \frac{1}{V}\int_{V}^{} \psi(x,t) \text{d}V=\frac{1}{V}\int_{V}^{} g(x) \text{d}V
\end{equation}
where the polynomial $g(x)$ of the level set is calculated based on a polynomial basis function as  
\begin{equation}
g(x)= < \psi>+\sum_{i}^{}c_{i}B_{i}(x)
\label{g_polynomial}
\end{equation}
where $c_i$ are the degrees of freedom used and $B_i$ are the basis functions calculated for each volume $V_P$ expressed via the orthogonal polynomial basis functions $L_{i}(x)$ using Legendre polynomials as
\begin{equation}
B_{i}(x)= L_{i}(x)-\frac{1}{V_{P}}\int_{V_{P}}^{} L_{i}(x) \text{d}V_{P}
\end{equation}
Eq.\eqref{hyperbolicPsiGauss} can be written with respect to the surface integrals over all faces $f$ of the cell as
\begin{equation}
\frac{\text{d}}{\text{d}t}{<\psi >}+\frac{1}{V} \sum_{N_{f}}^{}\int_{f}^{} F_{n_{f}} \text{d} f = 0
\label{hyperbolicFinal}
\end{equation}
where $F_{n_{f}}$ is the flux into the direction of the normal vector of each face $f$ from the total number of faces $N_f$. This is evaluated from the fluxes in the cells that share each face $f$, $\psi_L$, and $\psi_R$. We solve the Riemann problem interpolating both $\psi_L$, and $\psi_R$ with polynomials via  the expressions obtained through Eq.\eqref{g_polynomial}. 
More details for the implemented WENO stencil can be found in \cite{dumbser2007arbitrary,charest2015high}

Solving the transport equation for $\psi$ yields an updated value for level set function that might lose its smooth profile and might no longer remain a signed distance function.
Consequently, it needs to be re-initialised so that the Eikonal equation is satisfied, $|\nabla \psi|-1 =0$. This is a re-sharpening iterative procedure that uses as an initial guess the value obtained from Eq.~\eqref{psiDefinition} denoted with $\psi_0$.
The gradient magnitude in the Eikonal equation is calculated with the Godunov Hamiltonian $G$ according to
\begin{equation}
|\nabla \psi| \cong G(D^{-}_{x}\psi^n,D^{+}_{x}\psi^n,D^{-}_{y}\psi^n,D^{+}_{y}\psi^n,D^{-}_{z}\psi^n,D^{+}_{z}\psi^n)
\end{equation} 

The terms $D^{-}_{x}\psi^n, D^{+}_{x}\psi^n$,$D^{-}_{y}\psi^n, D^{+}_{y}\psi^n$, $D^{-}_{z}\psi^n, D^{+}_{z}\psi^n$ are the finite differences of $\psi$ at three directions x,y,z. For each $\zeta-$direction
\begin{linenomath*}
\begin{align}
D^{-}_{\zeta}\psi^n = \frac{\psi^n_{i} - \psi^n_{i-1}}{\Delta \zeta^-},  &   \hspace{1cm} D^{+}_{\zeta}\psi^n = \frac{\psi^n_{i+1} - \psi^n_{i}}{\Delta \zeta^+}
\end{align}
\end{linenomath*}
where $\Delta \zeta^-, \Delta \zeta^+$ are the distances of the cell centre $P$ with the cell centre of the upwind and downwind cells respectively. 
The Hamiltonian-Godunov term is then calculated as 
\begin{equation}
G= \sqrt{max(a^2_{x^-},a^2_{x^+}) +max(a^2_{y^-},a^2_{y^+}) + max(a^2_{z^-},a^2_{z^+}})
\end{equation}
Here, the terms $a_{x^-},a_{x^+}, a_{y^-},a_{y^+}, a_{z^-},a_{z^+}$ for each direction are calculated using the face normal vector of the level set. For instance at the x-direction, $a_{x^-},a_{x^+}$ are calculated using the unit vector $\mathbf{\Delta x}_{NP}$ of $\mathbf{x}_{NP}$. If $\psi <0$ and $\mathbf{x}_{NP} \cdot \mathbf{\Delta x}_{NP} < 0$ or $\psi >0$ and $\mathbf{x}_{NP} \cdot \mathbf{\Delta x}_{NP} > 0$ then  
\begin{equation}
a_{x} = min \left(\nabla^{\perp}_{f}\psi \cdot \mathbf{\Delta x}_{NP} \right)
\end{equation}
If $\psi <0$ and $\mathbf{x}_{NP} \cdot \mathbf{\Delta x}_{NP} > 0$ or $\psi >0$ and $\mathbf{x}_{NP} \cdot \mathbf{\Delta x}_{NP} < 0$ then 
\begin{equation}
a_{x} = max \left(\nabla^{\perp}_{f}\psi \cdot \mathbf{\Delta x}_{NP} \right)
\end{equation}  
where the normal gradient of the level set function $\nabla^{\perp}_{f}$ is calculated for all the faces $f$ based on the orientation of the normal at the face. In general
\begin{equation}
\nabla\psi^{\perp}_{f} = \alpha_{corr}(\psi_{P}-\psi_{n})/|\Delta \zeta|\hat{\zeta}+(\hat{\zeta}-\alpha_{corr}\Delta \zeta)\nabla(\psi)_{f}
\label{gradPsif}
\end{equation}
where $\nabla(\psi)_{f}$ is the gradient of $\psi$ that is interpolated at the face $f$ and $\alpha_{corr}$ is calculated from the angle $\theta$ between the cell centres and the normal face as $cos^{-1}(\theta)$.  
The following second order Runge-Kutta method is solved for a time-step $\tau$ \citep{Min2010}
\begin{linenomath*} 
\begin{align}
\begin{split}
&\psi_d^{'} =\psi_d^{n}-\Delta \tau \cdot sgn(\psi_0)\left[G(D^{-}_{x}\psi_d^n,D^{+}_{x}\psi_d^n,D^{-}_{y}     
\psi_d^n,D^{+}_{y}\psi_d^n,D^{-}_{z}\psi_d^n,D^{+}_{z}\psi_d^n)-1 \right] \\
&\psi_d^{''} =\psi_d^{'}-\Delta \tau \cdot sgn(\psi_0) \left[G(D^{-}_{x}\psi_d^{'},D^{+}_{x}\psi_d^{'},D^{-}_{y}   
\psi_d^{'},D^{+}_{y}\psi_d^{'},D^{-}_{z}\psi_d^{'},D^{+}_{z}\psi_d^{'})-1\right] \\
&\psi^{n+1}_d = \frac{\psi^{n}_d+\psi_d^{''}}{2}
\label{RK2}
\end{split}
\end{align}
\end{linenomath*}
Eq.~\eqref{RK2} is solved in a small narrow band surrounding the interface located at the $\psi$-0 level. This narrow band is considered to have a thickness $2\epsilon \Delta x$, where $\epsilon$ defines the layers of cells across which the re-initialisation is performed. Here, the values of $\epsilon=1.1-1.5$ were chosen \citep{Prosperetti2009}. The fictitious time step $\Delta \tau$ depends on the Courant number (CFL) which is 0.3-0.5 for the tests here, and the local cell-size, $\Delta \tau$ = CFL $\cdot$ min($\Delta \zeta ^-, \Delta \zeta ^+)$ \citep{SmerekaRusso2000,Dianat2017}. A small number of iterations was generally used here for obtaining $\nabla {\psi}_d$ that remained less than five. 
Alternatively, a different re-initialisation strategy could require spiting the cells at the interface employing a smooth calculation of the gradient of the level set function considering the neighbouring cells via an interpolation at the cell-faces. 
In \citep{Lyras2020} a second-order in space formulation was presented for both structured and unstructured polyhedral cells and can be also used in the present method for are simple and efficient calculation of the distance function across the interface. The latter will be presented in future studies.

The solutions of Eq.~\eqref{RK2} are distance functions and since sgn(0) = 0, they have the same zero-level as $\psi$. The new value of the level set function after the final re-initialisation step is denoted with $\tilde{\psi}_d$ e.g. $\tilde{\psi}_d=\psi^{n+1}_d$. 
The new level set value needs to provide a smoothed level set such that the $\int_{V}^{} H(\psi,t) \text{d}V$ is conserved. This means tha the new level set is now a signed distance function but it does not satisfy the condition in Eq.~\eqref{psiConservation} and thus will lead to mass loss \citep{Olsson2005}. 
In order to minimise this induced error, a correction $\hat{\psi}$ is added to $\tilde{\psi}_d$ as the proposed in \citep{Kees2011}. This second correction step is performed next for choosing the appropriate $\psi$ for a sharp profile that conserves mass.
The following reaction-diffusion equation is solved 
\begin{equation}
H(\psi_d + \hat{\psi}) = \alpha + \lambda \Delta \psi_d 
\label{massConservationCorr}
\end{equation}
where $\lambda$ is a parameter that is a function of the cell size and is generally taken here $\approx 0.5(\Omega)^{1/3}$, where $\Omega$ is the volume of the cell. Eq.~\eqref{massConservationCorr} is solved with an in-house implicit solver with the laplacian term using an unbounded, second-order conservative scheme using linear interpolation. The resulting $\hat{\psi}$ value leads to the final level set value  
\begin{equation}
\psi^{n+1} = \psi^{n+1}_d + \hat{\psi}
\end{equation}  
The new updated value satisfies the mass conservation since $H(\psi^{n+1}) = \alpha$ allowing the accurate curvature calculation which is included in the surface tension force in momentum equation in Eq.\eqref{momentumEqn} and Eq.\eqref{surfaceTensionEqn}. The latter step is extremely important since it avoids the assumption that the interface position is located at the iso-surface contour $\alpha$-0.5 (as in \citep{Kunkelmann2010} \citep{Albadawi2013}, \citep{Dianat2017}, \citep{Lyras2020}). Contrary to the previous works, the level set function is not reconstructed only from the advected volume fraction \citep{Albadawi2013}, \citep{Lyras2020} but neither is based on a specifically designed scheme that could require cell-searching algorithms that might require addtional runtime \citep{Lyras2020}, \citep{Hartmann2010}.
Contrary to other similar works \citep{Dianat2017} a Runge-Kutta scheme is used for the re-initialisation procedure that was previously shown to be second-order in the whole domain and more accurate than the forward Euler scheme \citep{Min2010}. This scheme, combined with the mass conservation reaction-diffusion step for the new value of $\psi$ aims at a low mass loss in Eq.\eqref{massLossEqn}. 
Apart of the parameters $\lambda,\epsilon$ no further tuning is required for this methodology.  

\subsection{Mass-conservation correction algorithm \& flux calculation}
\begin{figure}[h]
    \vspace{6pt}
    \centering
    \includegraphics[scale=0.28]{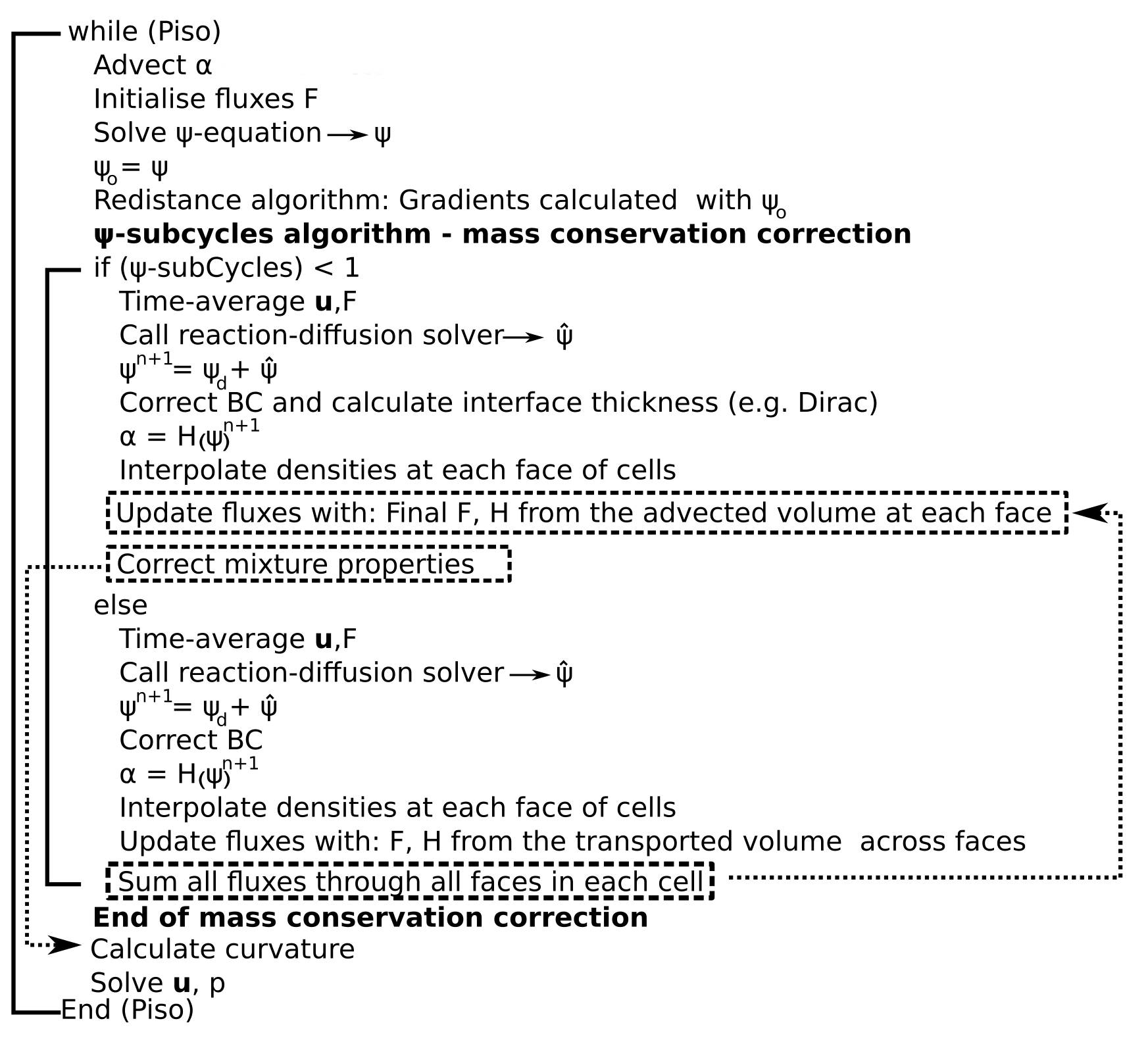}
       \centering
    \caption{Summary of the steps used for the level-set mass conservation algorithm.}
    \label{fig:algorithm}
\end{figure}

The mass correction is the most important step of the algorithm here and makes sure that the advected level set respects continuity equation. Since we use a finite volume approach it is vital to use an algorithm that successfully calculates the mass fluxes across all faces of the mesh cells and is applicable for different types of polyhedral meshes. 
In this study we developed an approach that deploys the smoothed volume fraction and calculates the fluxes in each face. An overview of the steps used in the solver to calculate the interface is shown in Fig.~\ref{fig:algorithm}. 

The mass-conservation algorithm is utilised in one or more cycles ($\psi$-sub-cycles). These are:
\begin{enumerate}
 \item First the fluxes are averaged based on the current and previous time-steps. Then the in-house reaction-diffusion solver is called that calculates the correction for level set, $\hat{\psi}$ based on the Eq.~\eqref{massConservationCorr}. For the evaluation of all the flux terms, this solver uses a third-order WENO (weighted essentially non-oscillatory) interpolation scheme which is implemented in the solver for the presented approach in this paper.
 \item The signed distance function is corrected calculating $\psi^{n+1}$ and updating boundary conditions.
 \item The smoothed volume fraction is calculated based on the Heaviside expression. 
 A smeared Heaviside function, $H$, is used for numerical robustness,\citep{Vukvcevic2016} defined as
 \begin{equation}
  H(\psi)
  = \begin{cases}
  0 & \text{if  $\psi < - \epsilon$}  \\
  \frac{1}{2}\left[ 1 + tanh (\frac{\psi}{\sqrt 2\epsilon}) \right] & \text{if $|\psi|\leq \epsilon$} \\
  1 & \text{if  $\psi > \epsilon$}  
  \end{cases}
 \end{equation} 
This new value is used to interpolate the densities of the fluids $\rho_{1}, \rho_{2}$ and their difference $(\Delta \rho)_{f}$ at the cell faces in all cells.
 \item At this stage, the level set function is a signed distance function which satisfies Eq.~\eqref{psiConservation}  but the fluxes need to be re-calculated. The volumetric fluxes $F$ across mesh faces $f$ are calculated based on the new smoothed volume fraction and interpolated densities.  
If this is not the final $\psi$-sub-cycle then, all the mass fluxes are summed up per cell. At the final $\psi$-sub-cycle this sum is used as the mass flux in all faces. In general one $\psi$-sub-cycle is enough for good accuracy but the user can use more for improving mass correction if the problems demand it.  
 \item Using the new mass flux the mixture properties are corrected and the new curvature for the interface is obtained. This curvature is used in the momentum equation for the surface tension force. 
\end{enumerate}

\subsection{Incompressible two-phase flow equations}
The governing incompressible Navier-Stokes equations for two-phase flow with surface tension are given by

\begin{equation}
\frac{ \partial u_{j}}{\partial x_{j}}= 0 
\label{massEqn}
\end{equation}

\begin{equation}
\rho \left[ \frac{\partial u_{i}}{\partial t}+ u_{j}\frac{\partial u_{i}}{\partial x_{j}} \right] = -\frac{\partial p}{\partial x_{i}}+ \frac{\partial \tau_{ij}}{\partial x_{j}} + F_{\sigma}
\label{momentumEqn}
\end{equation}
where $u_{i}(x, t)$ represents the $i$-th component of the fluid velocity at a point in space, $x_{i}$, and time, $t$. Also $p(x, t)$ represents the static pressure, $\tau_{ij}(x, t)$, the viscous (or deviatoric) stresses, and $\rho$ the fluid density. The deviatoric stress tensor is given by $\tau_{ij}=2\mu S_{ij}$ ,where $S_{ij} = 0.5(\frac{\partial u_{i}}{\partial x_{j}} + \frac{\partial u_{j}}{\partial x_{i}})$.
Once the volume fraction is advected, the level set equation is solved and is corrected using Eq.\eqref{RK2} obtaining $\nabla{\psi}_{d}$. The reaction-diffusion equation is solved obtaining the new level set function $\psi^{n+1}$. A correction for the volume fraction is performed based on $\psi_{d}$ such that 
\begin{equation}
\alpha_{H}=H(\psi^{n+1})=\alpha
\label{alphaH}
\end{equation}
where $\alpha_{H}$ is the smoothed volume fraction from a Heaviside expression that depends on the value of $\psi$. This step is very important since it links level set with the calculated volume fraction in each computational cell. 

The newly corrected volume fraction $\alpha_{H}$ is then used for calculating the fluxes across the faces of each cell and is used into the momentum equation. The mixture properties are calculated with the one-fluid approach, where density and viscosity of the pseudo-fluid are calculated as \citep{Tryggvason2011}. 
\begin{equation}
\rho = \rho_1 H + \rho_2 (1-H) 
\end{equation}
\begin{equation}
\mu = \mu_1 H + \mu_2 (1-H)
\end{equation}
The surface tension force acting on the interface is calculated as \citep{Brackbill1992}
\begin{equation}
F_{\sigma}=\sigma \kappa \delta(\psi)\nabla \psi 
\end{equation}
The Dirac function $\delta()$ is used for smoothing the surface tension force effect at the interface. Depending on the interface thickness $\delta()$ is evaluated from \citep{Albadawi2013,Lyras2020}
\begin{equation}
\delta(\psi)
= \begin{cases}
0 & \text{if  $|\psi| >  \epsilon$}  \\
\frac{1}{2 \epsilon}\left[ 1+cos\left( \frac{\pi \psi}{\epsilon} \right)\right] & \text{if $|\psi| \leq \epsilon$} 
\end{cases}
\end{equation}
The curvature defined as $\kappa= - \nabla \cdot \psi_{\kappa}$, where $\psi_{\kappa}=\nabla \psi /|\nabla \psi|$ and the gradient can be discretised (since $\psi$ is a continuous function) and the surface tension force is calculated at the cell faces using
\begin{equation}
F_{\sigma} = \left(\sigma \mathbf{\kappa} \right)_f \delta_f(\psi)\nabla^{\perp}_{f} \psi
\label{surfaceTensionEqn} 
\end{equation} 
where $\left(\sigma \mathbf{\kappa} \right)_f$ and $\delta_f$ are interpolated at the faces of the cell.

\section{Numerical tests and discussions}
In this section, the proposed methodology is evaluated for capturing the interface between two immiscible fluids with some widely used advection benchmark tests. The rotating disc and sphere tests were performed to assess the capability of the method to capture the interface of rotating filaments under a non-uniform velocity field. Zalesak's slotted disc test is presented next, and the motion of a rising bubble in two and three dimensions is presented as an example of a flow where surface tension force plays an important role. 
In order to quantify the accuracy of the numerical results, the shape preservation error $E_{\alpha}$ is calculated, defined as
The error 
\begin{equation}
E_{\alpha}=\frac{\sum_{i}^{ }V_i|\alpha_i-\alpha_{i,exact}|}{\sum_{i}^{}V_i\alpha_{i,exact}}
\end{equation}
the sum considers all cells in the domain with numerical solution $\alpha_i$. The exact solution $\alpha_{i,exact}$ is obtained from the position of the disc at t=0s. 
The advection error is also quantified using
\begin{equation}
L_{1}(\alpha)=\sum_{i}^{ }V_i|\alpha_i-\alpha_{i,exact}|
\end{equation}

\subsection{Two-dimensional vortex deformation transport}
\begin{figure}[H]
 \vspace{6pt}
\centering
\begin{subfigure}{.32\textwidth}
  \centering
  \includegraphics[width=0.8\linewidth]{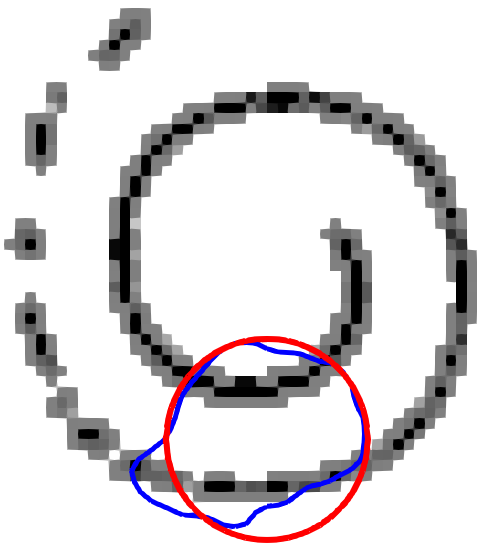}
  \caption{Coarse quadrilateral}
  \label{fig:sub1}
\end{subfigure}%
\begin{subfigure}{.32\textwidth}
  \centering
  \includegraphics[width=0.8\linewidth]{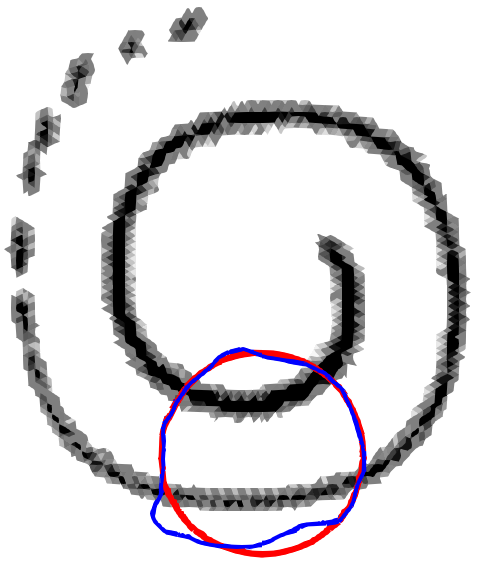}
  \caption{Coarse triangular}
  \label{fig:sub2}
\end{subfigure}
\begin{subfigure}{.32\textwidth}
  \centering
  \includegraphics[width=0.8\linewidth]{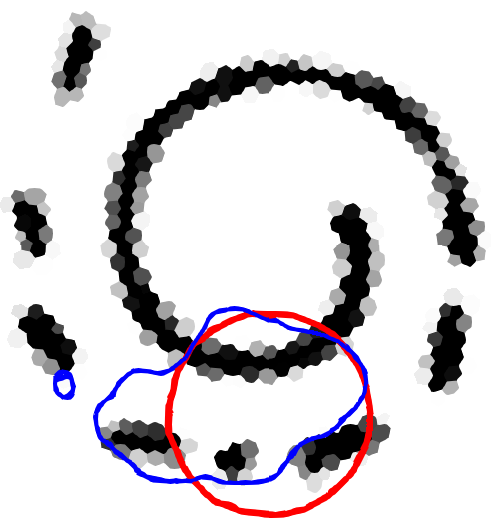}
  \caption{Coarse polyhedral}
  \label{fig:sub2}
\end{subfigure}
\begin{subfigure}{.32\textwidth}
  \centering
  \includegraphics[width=0.8\linewidth]{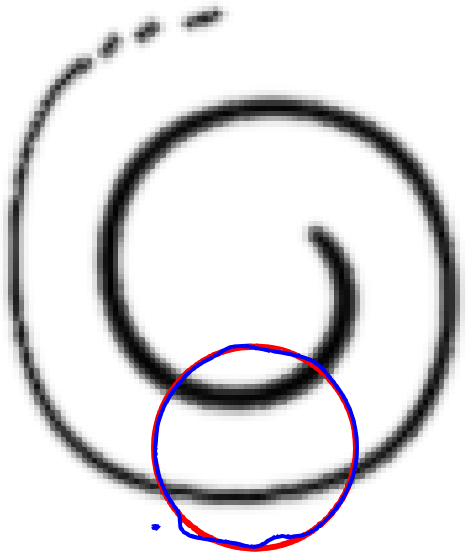}
  \caption{Medium quadrilateral}
  \label{fig:sub3}
\end{subfigure}%
\begin{subfigure}{.32\textwidth}
  \centering
  \includegraphics[width=0.8\linewidth]{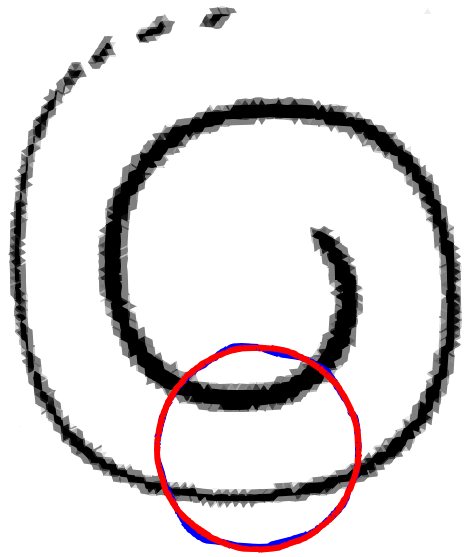}
  \caption{Medium triangular}
  \label{fig:sub4}
\end{subfigure}%
\begin{subfigure}{.32\textwidth}
  \centering
  \includegraphics[width=0.8\linewidth]{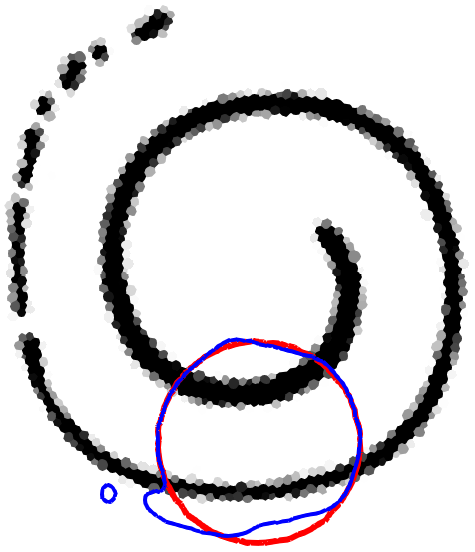}
  \caption{Medium polyhedral}
  \label{fig:sub2}
\end{subfigure}
\begin{subfigure}{.32\textwidth}
  \centering
  \includegraphics[width=0.8\linewidth]{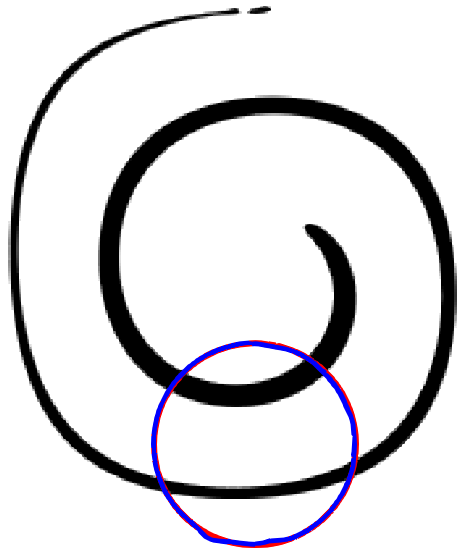}
  \caption{Fine quadrilateral}
  \label{fig:sub5}
\end{subfigure}%
\begin{subfigure}{.32\textwidth}
  \centering
  \includegraphics[width=0.8\linewidth]{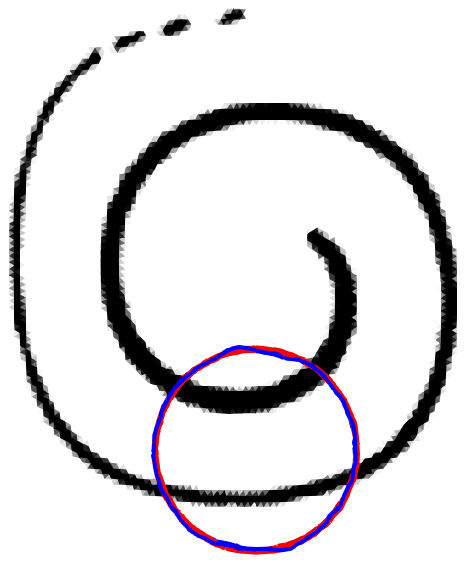}
  \caption{Fine triangular}
  \label{fig:sub6}
\end{subfigure}%
\begin{subfigure}{.32\textwidth}
  \centering
  \includegraphics[width=0.8\linewidth]{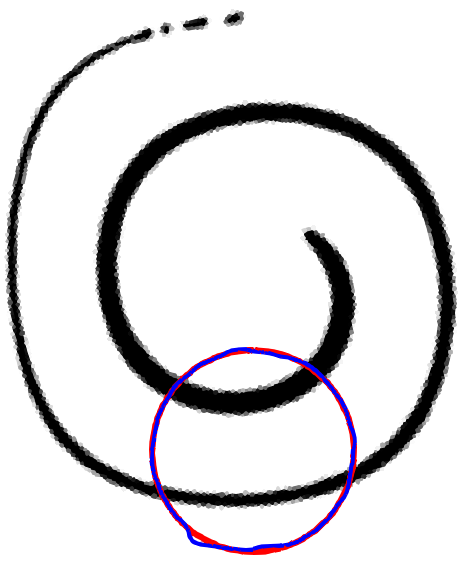}
  \caption{Fine polyhedral}
  \label{fig:sub6}
\end{subfigure}%
\caption{Two-dimensional rotating disc test for the presented method at $t=T/2$. The smoothed volume fraction H, the initial (light purple line) at $t=0$ and final position of the zero-level set iso-surface (blue line) at $t=T$ are indicated.}
\label{fig:2DrotDisk_LS_performance}
\end{figure}

In this test an initially still disc filled with liquid starts to rotate under the influence of a velocity field that changes through time. The disc progressively deforms and streches until it becomes a ligament. After reaching its maximum deformation, the flow reverses, which causes the ligament to return back to its original shape. This classical numerical test of \citep{LeVeque1996} demonstrates the capability of the numerical method to capture highly deformed interfaces. The disc is placed in a two-dimensional domain $[0,1]^2$ and at t=0s has radius $R$ with the disc centre at $(x,y)=(0.5,0.75)$. The velocity field is given by 
\begin{equation}
u(x,y,t)=-sin^2(\pi x)sin(2\pi y)cos\left( \frac{\pi t}{T} \right)
\label{1}
\end{equation}
\begin{equation}
v(x,y,t)=sin(2\pi x)sin^2(\pi y)cos\left( \frac{\pi t}{T} \right)
\label{2}
\end{equation}
For this test, the simulation lasted for one period, T=8 and the Courant number was 0.3. A fixed Eulerian grid is used for all cases. The total simulation time was $8s$.
The interface is initially a circle which is distorted and stretched through time with a maximum deformation at t=T/2. At that point the vortex has become a spiral-like ligament with a thin tail. 
Increasing t, at $t \in \left[0, T/2 \right]$ deforms the rotating vortex and the thickness of the spiral becomes smaller than the grid size.
Thus, the spiral breaks up into smaller pieces, with droplets generated due to the under-resolved liquid fraction field. Consequently, the interface loses its smooth shape. 
For $t \in \left[T/2, T \right]$ the flow reverses and the spiral is restored back to its originally circular shape.
The loss in geometrical information, is also evident during the reversed flow, and the resulting restored disc has no longer the ideal circular shape at t=T.
Different meshes were tested for structured, unstructured and polyhedral meshes with three different grid resolutions indicated as coarse, medium and fine. For the structured meshes the resolutions used were 32 $\times$ 32, 64 $\times$ 64 and 128 $\times$ 128. For the unstructured and polyhedral meshes, the resolutions were similar and are listed in Table~\ref{table:2DRotatingDisk_Ea}. 
The numerical results for the different grids at t=0, T/2, T are shown in Fig.~\ref{fig:2DrotDisk_LS_performance} for the smoothed volume fraction at t=T/2 and 0.5-contour of the field at t=0, T. 

The mesh resolution plays an important role for capturing the interface and in all cases it plays a major role to the ligament fragmentation due to the strong deformation. The droplet formation is more evident for coarse meshes at t=T/2, with a significant improvement for the finer meshes where the local cell size becomes closer to the local thickness of the stretched ligament.   

The results for shape preservation also illustrate how the mesh resolution influences the 
in Table~\ref{table:2DRotatingDisk_Ea}. Overall, the error was reasonably low regardless the mesh type. The re-initialisation step for the level set function considered the Runge-Kutta scheme which has shown third-order accuracy close to the interface and that it might be at least second-order for the rest of the domain \citep{Min2010}. Calculating gradients across the interface separating two fluids poses challenges which can result in the mass error introduced before solving the reaction-diffusion equation for obtaining the smoothed volume fraction. Here, an central-difference scheme in Eq.\ref{gradPsif} is employed for calculating the normal gradient of the level set function with an extra term for non-orthogonal meshes. The non-orthogonal correction with the explicit non-orthogonal term becomes important when $\theta$ was more than 30$^o$ and the meshes here had generally non-orthogonality less than 55$^o$.  
 
The results for the structured mesh are also compared with other numerical methods for either VOF or coupled methods with level set in Table~\ref{table:2DRotatingDisk_L1}. The results here are reasonably close to the errors reported from the coupled Tangent of Hyperbola Interface Capturing
(THINC) and level set methods in ~\citep{Yokoi2007,Qian2018}. The CLSVOF method of ~\citep{Qian2018} using a  surface representation based on a first/forth order polynomial, THINC-LS (P1/P4) has previously demonstrated excellent results for mass conservation and interface representation, offering significant improvements compared to other geometric methods ~\citep{Qian2018}. 
The error obtained with the presented method here were close or lower than other CLSVOF methods such as the ones in ~\citep{Jemison2013,Ningegowda2014,Singh2018} and other VOF methods such as ~\citep{Rider1998}. Similarly, the results were lower to the ones obtained with the PLIC method of ~\citep{Aulisa2003}. The method here does not employ an explicit interface reconstruction step or other surface interpolation in space that might be dependent on the mesh topology. Thus, it has a simple level set implementation that can be naturally implemented in various types of meshes. 
The approach here generally shows an accuracy which is close or better than other PLIC-VOF methods and coupled level-set-VOF methods, including the one in ILSVOF \citep{Lyras2020}. We improve the accuracy of the ILSVOF by using a fundamentally different algorithm of the level set advection: the ILSVOF method in \citep{Lyras2020} does not solve any equation for level-set and uses a specific second-order re-initialisation scheme to compensate for interface smearing. 
On the other hand, here we solve the level set equation and we do not locate the interface as the 0.5-contour of the volume fraction as in ILSVOF, but solve an extra equation to guarantee that the interface representation recovers mass conservation. The results show that with this approach, one might generally be able to avoid using dedicated re-initialisation schemes for re-sharpening level set as in \citep{Hartmann2010,Lyras2020}. 

\begin{figure}[H]
    \vspace{6pt}
    \centering
    \includegraphics[scale=0.16]{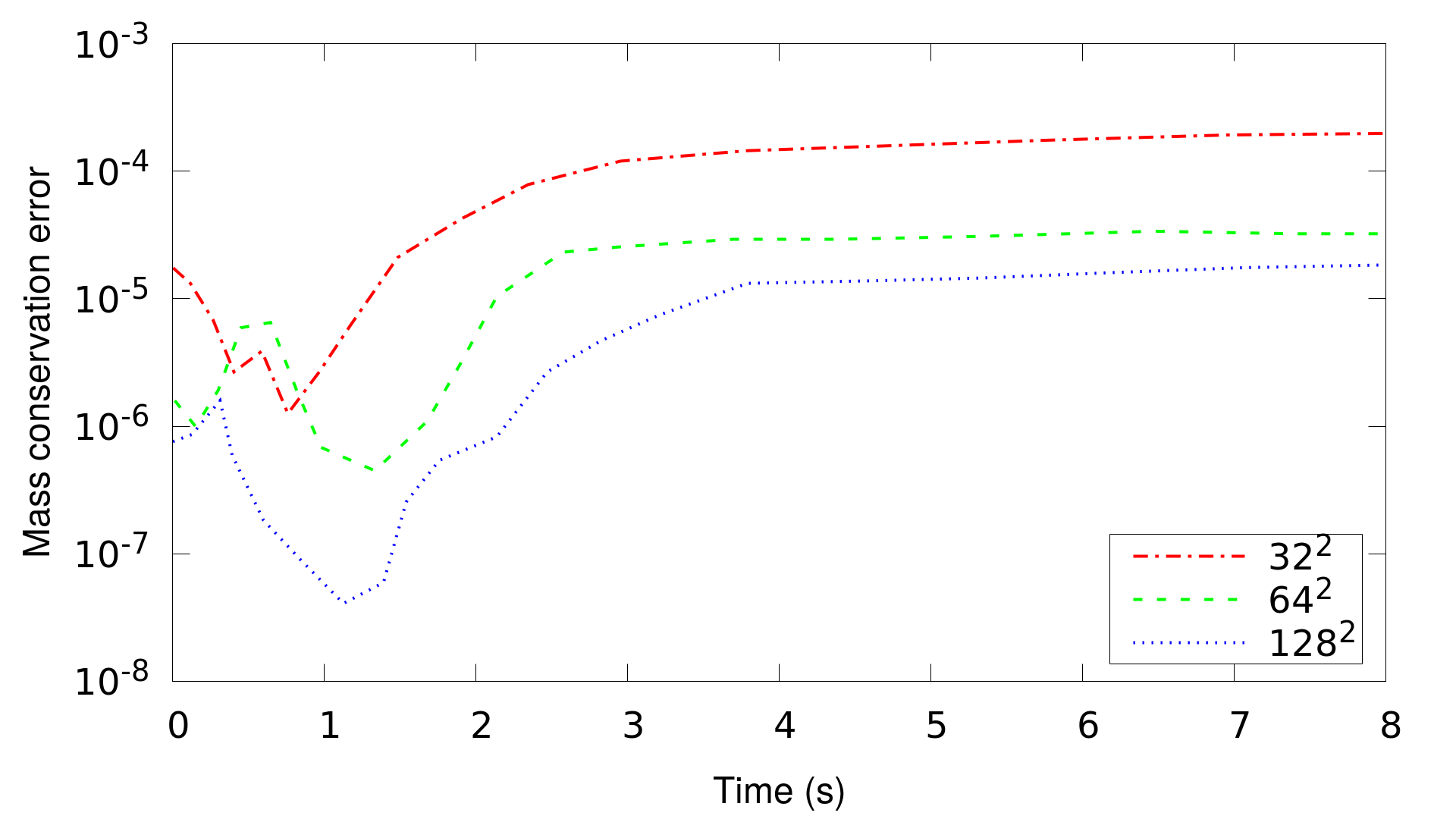}
       \centering
    \caption{Mass conservation error for the two-dimensional rotating disc case for the three levels of refinement, $32^{2}$, $64^{2}$, $128^{2}$.}
    \label{fig:2D_rotating_sphere_mass_conservation}
\end{figure}

The mass conservation error time history is shown in Fig.~\ref{fig:2D_rotating_sphere_mass_conservation}. 
Overall, the proposed method demonstrated reasonably low mass conservation error for the structured mesh tests. Mass conservation is based on the Eq.\ref{massConservationCorr} with the parameter $\lambda$ being  here equal to $50\%$ of the local mesh size.

\begin{table}
\caption{$E_{\alpha}$ calculated using different meshes for the 2D rotating disc case.}
\centering
\label{table:2DRotatingDisk_Ea}
\begin{tabular}{l l l}
\hline
Mesh & Resolution & $E_{\alpha}$ \\
\hline
 & $64^2$ & -2.11$\cdot 10^{-7}$\\

Structured & $128^2$ & -3.69$\cdot 10^{-8}$\\

 & $256^2$  & -8.13$\cdot 10^{-9}$\\
 
 \midrule
 
& $4112$ & -5.19$\cdot 10^{-6}$\\

Unstructured & $16010$ & -5.87$\cdot 10^{-7}$\\

 & $64020$  & 8.13$\cdot 10^{-8}$\\
 
 \midrule

& $4062$ & 4.21$\cdot 10^{-5}$\\

Polyhedral &$16033$ & 8.43$\cdot 10^{-6}$\\

 &$64012$ & -3.16$\cdot 10^{-6}$\\
 
\hline
\end{tabular}
\end{table}

\begin{table}
\caption{Comparisons of the methods using quadrilateral meshes for the two-dimensional rotating disc case. The first order norm $L_{1}(\alpha)$ is calculated for the three different meshes.}
\centering
\label{table:2DRotatingDisk_L1}
\begin{tabular}{l l l l}
\hline
Authors & $32^{2}$ & $64^{2}$ & $128^{2}$ \\
\hline

DS-CLSMOF ~\citep{Jemison2013} & 2.92$\times 10^{-2}$ & 5.51$\times 10^{-3}$ & 1.37$\times 10^{-3}$  \\

PLIC ~\citep{Aulisa2003} & 2.53$\times 10^{-2}$ & 2.78$\times 10^{-3}$ & 4.8$\times 10^{-4}$  \\

THINC/WLIC ~\citep{Yokoi2007} & 4.16$\times 10^{-2}$ & 1.61$\times 10^{-2}$ & 3.56$\times 10^{-3}$  \\

THINC-LS (P1) ~\citep{Qian2018} & 6.71$\times 10^{-2}$ & 1.53$\times 10^{-2}$ & 2.27$\times 10^{-3}$  \\

THINC-LS (P4) ~\citep{Qian2018} & 2.85$\times 10^{-2}$ & 3.39$\times 10^{-3}$ & 6.79$\times 10^{-4}$  \\

Rider-Kothe/Puckett ~\citep{Rider1998} & 4.78$\times 10^{-2}$ & 6.96$\times 10^{-3}$ &  1.44$\times 10^{-3}$ \\

InterFOAM  ~\citep{Deshpande2012} & 1.026$\times 10^{-1}$ & 3.65$\times 10^{-2}$ & 2.32$\times 10^{-2}$  \\

CLSVOF in ~\citep{Ningegowda2014} & - & - & 3.9236$\times 10^{-3}$  \\

CLSVOF in ~\citep{Singh2018} & - & - & 1.93$\times 10^{-3}$  \\

ILSVOF in ~\citep{Lyras2020} & 4.19$\times 10^{-2}$ & 1.43$\times 10^{-3}$ & 8.36$\times 10^{-4}$  \\

Presented method & 1.27$\times 10^{-2}$ & 1.14$\times 10^{-3}$ & 2.33$\times 10^{-4}$  \\

\hline
\end{tabular}
\end{table}

\subsection{Three-dimensional vortex deformation}
\begin{figure}[H]
 \vspace{2pt}
\centering
\begin{subfigure}{.28\textwidth}
  \centering
  \includegraphics[width=1\linewidth]{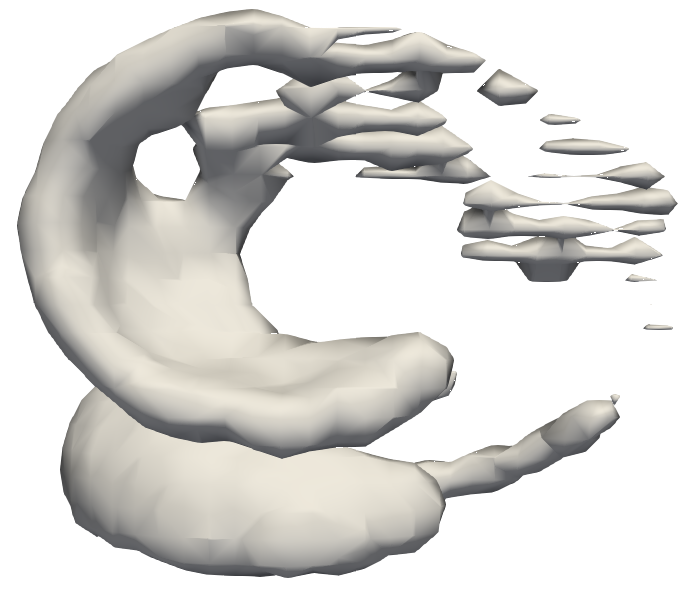}
  \label{fig:sub4}
\end{subfigure}%
\begin{subfigure}{.28\textwidth}
  \centering
  \includegraphics[width=1\linewidth]{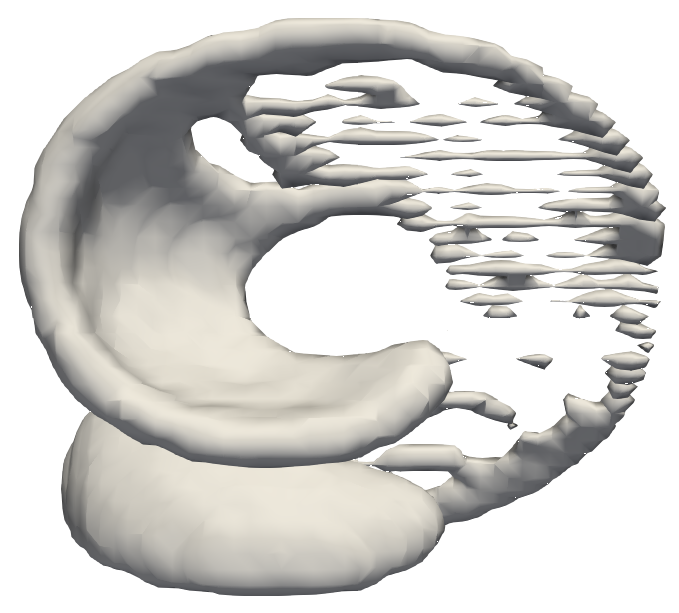}
  \label{fig:sub5}
\end{subfigure}
\begin{subfigure}{.28\textwidth}
  \centering
  \includegraphics[width=1\linewidth]{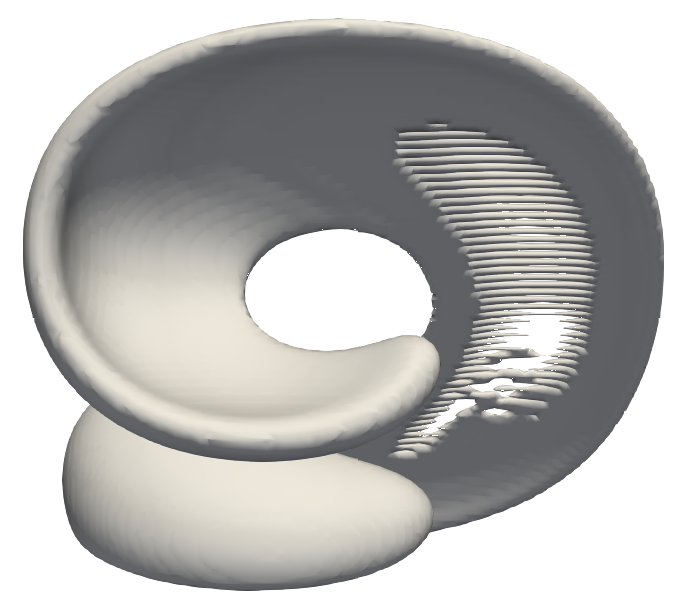}
  \label{fig:sub6}
\end{subfigure}%

\begin{subfigure}{.26\textwidth}
  \centering
  \includegraphics[width=1\linewidth]{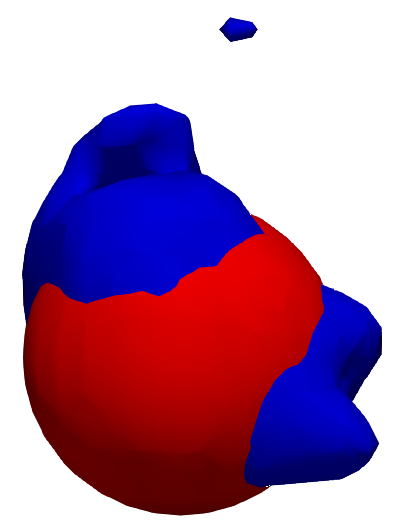}
  \caption{$32^{3}$}
  \label{fig:sub4}
\end{subfigure}%
\begin{subfigure}{.33\textwidth}
  \centering
  \includegraphics[width=1\linewidth]{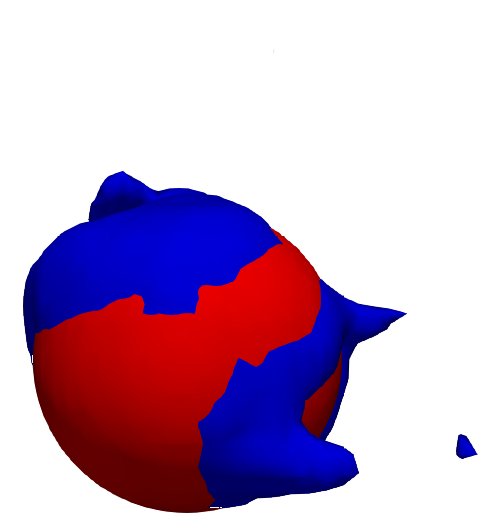}
  \caption{$64^{3}$}
  \label{fig:sub5}
\end{subfigure}
\begin{subfigure}{.29\textwidth}
  \centering
  \includegraphics[width=1\linewidth]{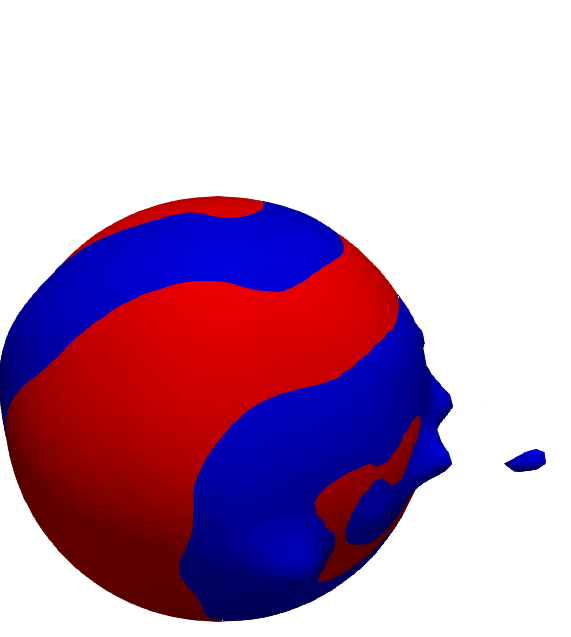}
  \caption{$128^{3}$}
  \label{fig:sub6}
\end{subfigure}%
\caption{Three-dimensional rotating sphere in a non-uniform flow test for various levels of structured mesh. The 0.5-iso-surface of H obtained with the presented method at $t=T/2$ (top) and at $t=0$ with red and at $t=T$ with blue (bottom) are shown.}
\label{fig:3DrotSphere_structured}
\end{figure}

The methodology is assessed next for three-dimensional interfaces undergoing strong deformations performing the test proposed in \citep{LeVeque1996}. A sphere with radius 0.15 is centred at (0.35, 0.35, 0.35) inside a domain $[0, 1]^3$ filled with gas. The liquid sphere is transported by a time-dependent velocity field given by
\begin{linenomath*} 
\begin{align}
\begin{split}
u(x,y,z,t)= 2sin^2(\pi x)sin(2\pi y)sin(2\pi z)cos\left(\frac{\pi t}{T}\right) \\
v(x,y,z,t)= -sin(2\pi x)sin^2(\pi y)sin(2\pi z)cos\left(\frac{\pi t}{T}\right) \\
w(x,y,z,t)= -sin(2\pi x)sin(2\pi y)sin^2(\pi z)cos\left(\frac{\pi t}{T}\right)
\end{split}
\end{align} 
\end{linenomath*} 
The period of the simulation was $T=3s$ and the sphere was stretched until t=T/2s when the flow reversed unstretching the sphere until t=T.
Three different levels of refinement were used for the tests for structured and unstructured meshes. The grid resolutions indicated as coarse, medium and fine were 32$^3$, 64$^3$ and 128$^3$. For the unstructured meshes, the resolutions were similar and are listed in Table~\ref{table:3DRotatingSphere_Ea}. 
Fig.~\ref{fig:3DrotSphere_structured} shows the smoothed field $H$ at maximum stretching $t=T/2$ and at the initial and final positions of the sphere, $t=T$. 
At the maximum stretching the 3D sphere resembles a thin sheet that is deformed in all three directions. During the rotation the liquid sheet breaks up into larger fragments which is more evident in the coarse meshes for both hexahedral and tetrahedral meshes, with voids of gas inside the rotating vortex.  
Increasing the mesh resolution, increases the accuracy of interface capturing for the liquid sheet with less voids present. 
The discrepancies that occur at the interface at $t=T$ for the medium and fine meshes due to the additional break up of the ligaments, are less evident since the mesh resolution becomes approximately equal to the droplet size.
The error $E_{\alpha}$ is shown in Table~\ref{table:3DRotatingSphere_Ea} for the different meshes.
The error is lower for the structured grids in general. Calculating the gradient of $\psi$ for tetrahedral meshes is challenging and requires high order schemes when solving the level set equation. Furthermore, some algorithms for correcting the gradient in the vicinity of the interface for tetrahedral meshes can be used to properly calculate the re-initialised level set function \citep{Lyras2020}.  

The results for $L_{1}(\alpha)$ are also shown in Table~\ref{table:3DRotatingDisk_L1} for comparisons with other numerical methods using hexahedral meshes. 
The error is significantly lower than the results obtained in \citep{Deshpande2012} using the solver interFoam of \citep{Weller2008}. The later uses is an algebraic VOF method that has been previously shown to suffer from diffusion in the 3D rotating disc test \citep{Roenby2016}. The solver interFoam is also limited to smaller Courant numbers for stability reasons (no more than 0.1 is preferred). 
  
The results were better for the different grids compared to the RK-3D method of ~\citep{Hernandez2008} using PLIC. 
Compared to the CLSVOF method of ~\citep{Qian2018}, the error was lower than the ones obtained with THINC-LS (P1) and the THINC-LS (P4). Overall, the method is proven to be competitive to the other geometrical VOF methods having the advantage of no interface reconstruction regardless the mesh and shows a significant improvement in accuracy with 50-100$\%$ lower error for most cases with the presented formulation.  
 
\begin{figure}[H]
 \vspace{2pt}
\centering
\begin{subfigure}{.28\textwidth}
  \centering
  \includegraphics[width=1\linewidth]{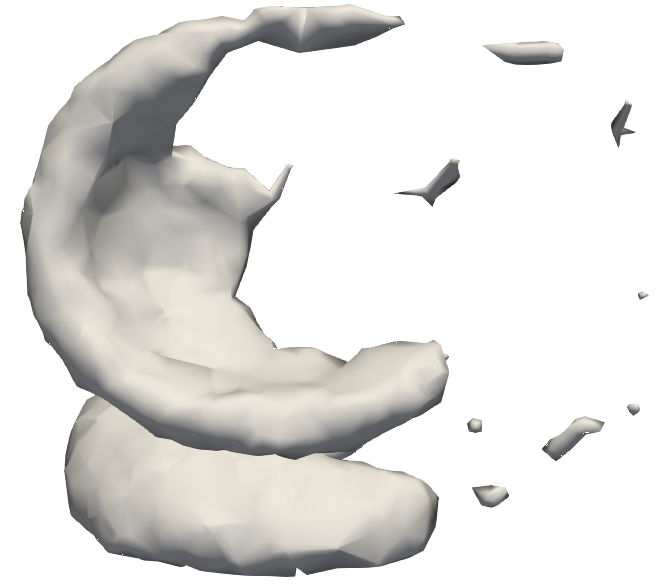}
  \label{fig:sub4}
\end{subfigure}%
\begin{subfigure}{.28\textwidth}
  \centering
  \includegraphics[width=1\linewidth]{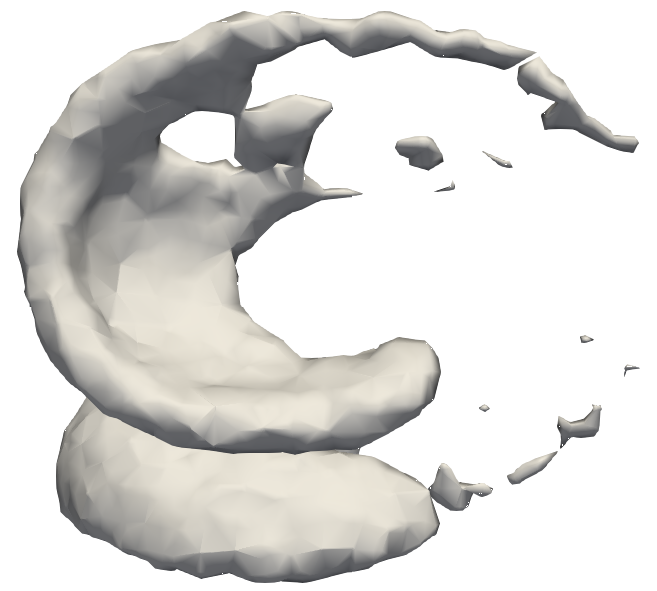}
  \label{fig:sub5}
\end{subfigure}
\begin{subfigure}{.28\textwidth}
  \centering
  \includegraphics[width=1\linewidth]{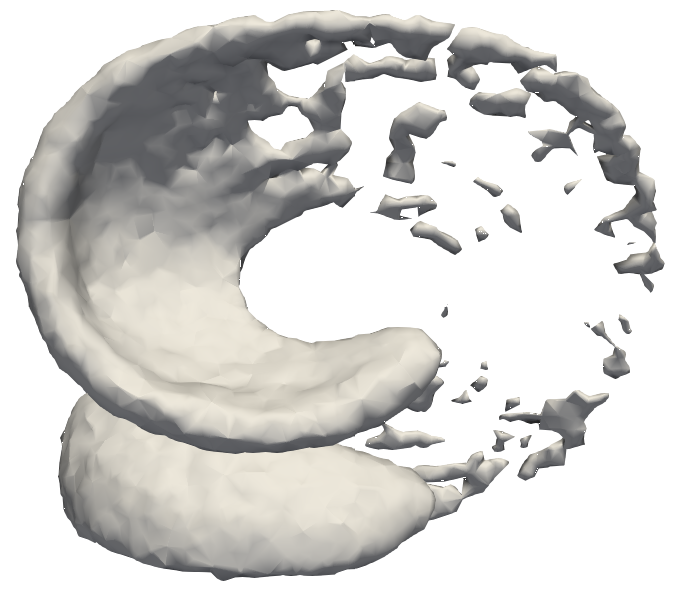}
  \label{fig:sub6}
\end{subfigure}%

\begin{subfigure}{.26\textwidth}
  \centering
  \includegraphics[width=1\linewidth]{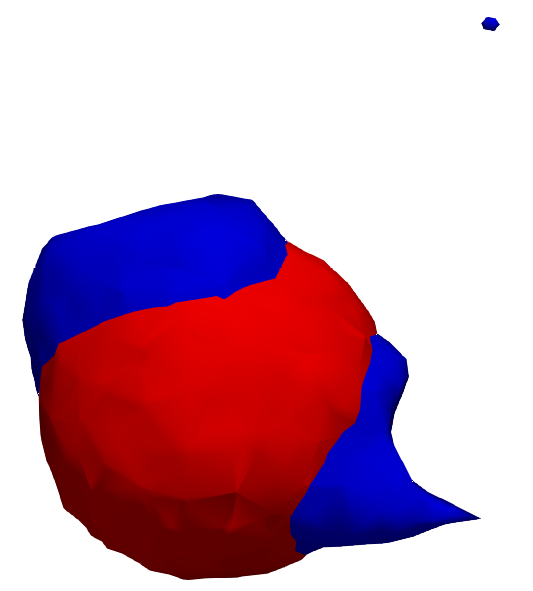}
  \caption{$81547$}
  \label{fig:sub4}
\end{subfigure}%
\begin{subfigure}{.33\textwidth}
  \centering
  \includegraphics[width=1\linewidth]{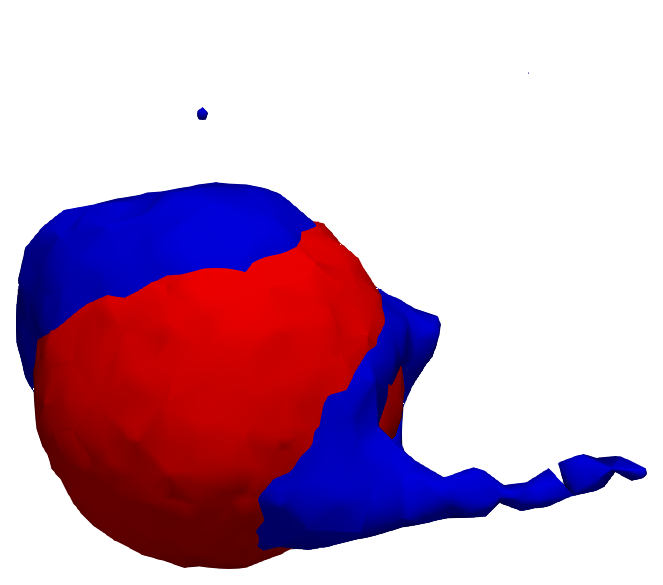}
  \caption{$226733$}
  \label{fig:sub5}
\end{subfigure}
\begin{subfigure}{.3\textwidth}
  \centering
  \includegraphics[width=1\linewidth]{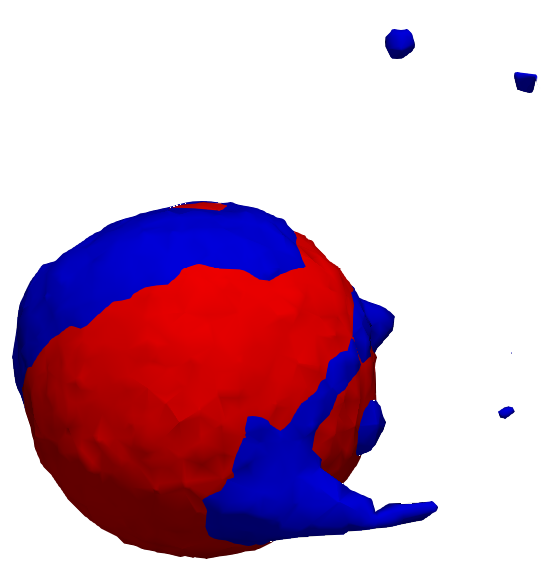}
  \caption{$1200182$}
  \label{fig:sub6}
\end{subfigure}%
\caption{Three-dimensional rotating sphere in a non-uniform flow test for various levels of unstructured mesh. The 0.5-iso-surface of H obtained with the presented method at $t=T/2$ (top) and at $t=0$ with red and at $t=T$ with blue (bottom) are shown.}
\label{fig:3DrotSphere_unstructured}
\end{figure}

\begin{table}
\caption{$E_{\alpha}$ error calculated using different meshes for the 3D rotating sphere test.}
\centering
\label{table:3DRotatingSphere_Ea}
\begin{tabular}{l l l}
\hline
Mesh & Resolution & $E_{\alpha}$ \\
\hline
 & $32^3$ & -2.21$\cdot 10^{-6}$\\

Structured & $64^3$ & -3.33$\cdot 10^{-7}$\\

 & $128^3$  & -1.27$\cdot 10^{-7}$\\
 
 \midrule
 & 81547 & -5.43$\cdot 10^{-4}$\\

Unstructured & 226733 & -7.22$\times 10^{-5}$\\

 & 1200182 & -2.66$\cdot 10^{-5}$\\
 
\hline
\end{tabular}
\end{table}
             
\begin{table}
\caption{$L_{1}(\alpha)$ error norm for structured meshes for the three-dimensional rotating sphere case and comparison with other methods.}
\centering
\label{table:3DRotatingDisk_L1}
\begin{tabular}{l l l l}
\hline
Authors & $32^{3}$ & $64^{3}$ & $128^{3}$ \\
\hline
RK-3D using PLIC ~\citep{Hernandez2008} & 7.85$\times10^{-3}$ & 2.75$\times10^{-3}$ & 7.41$\times 10^{-4}$  \\

THINC-LS (P1) ~\citep{Qian2018} & 7.18$\times 10^{-3}$ & 2.34$\times 10^{-3}$ & 6.14$\times 10^{-4}$  \\

THINC-LS (P4) ~\citep{Qian2018} & 5.54$\times 10^{-3}$ & 1.57$\times 10^{-3}$ & 3.79$\times 10^{-4}$  \\

interFoam ~\citep{Deshpande2012} & 9.95$\times 10^{-3}$ & 4.78$\times 10^{-3}$ & 2.03$\times 10^{-3}$  \\
     
ILSVOF in ~\citep{Lyras2020} & 8.89$\times 10^{-2}$ & 2.96$\times 10^{-3}$ & 8.06$\times 10^{-4}$  \\
         
Presented method & 1.16$\times 10^{-3}$ & 4.72$\times 10^{-4}$ & 8.68$\times 10^{-5}$  \\

\hline
\end{tabular}
\end{table}

\subsection{Zalesak's disc}
\begin{figure}[h]
    \vspace{6pt}
    \centering
    \includegraphics[scale=0.20]{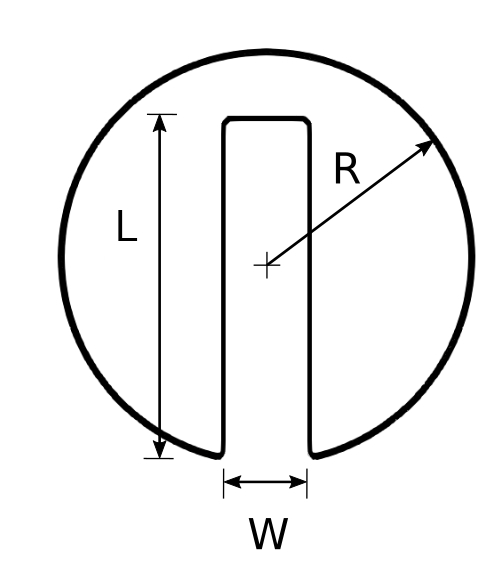}
       \centering
    \caption{Domain used for Zalesak's disc test case.}
    \label{fig:zalesak_domain}
\end{figure}

\begin{figure}[h]
    \vspace{6pt}
    \centering
    \includegraphics[scale=0.15]{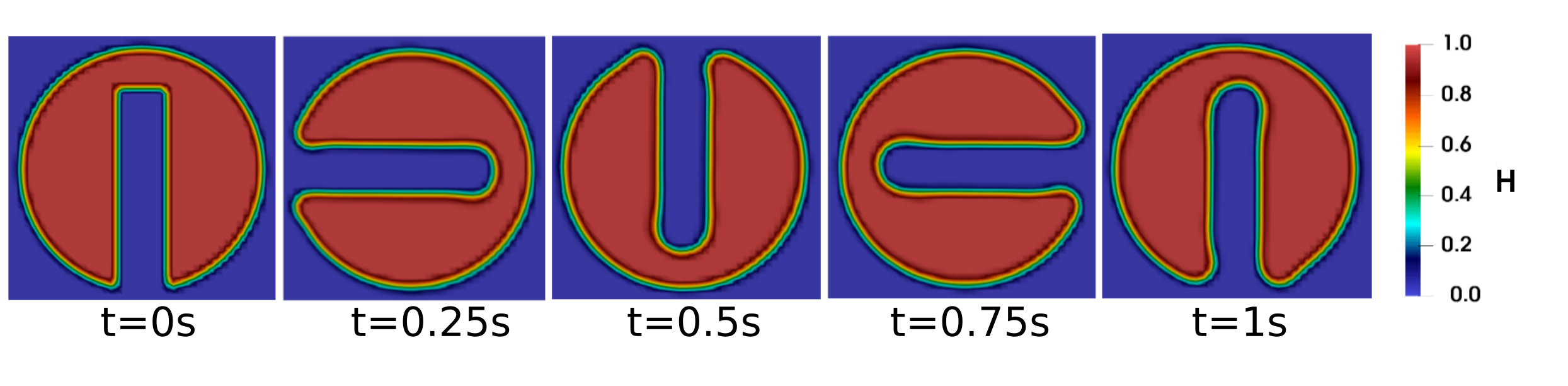}
       \centering
    \caption{Calculated smoothed level set for Zalesak’s disc for one rotation.}
    \label{fig:zalesak_one_rotation}
\end{figure}

\begin{figure}[H]
 \vspace{6pt}
\centering
\begin{subfigure}{.4287\textwidth}
  \centering
  \includegraphics[width=1.0\linewidth]{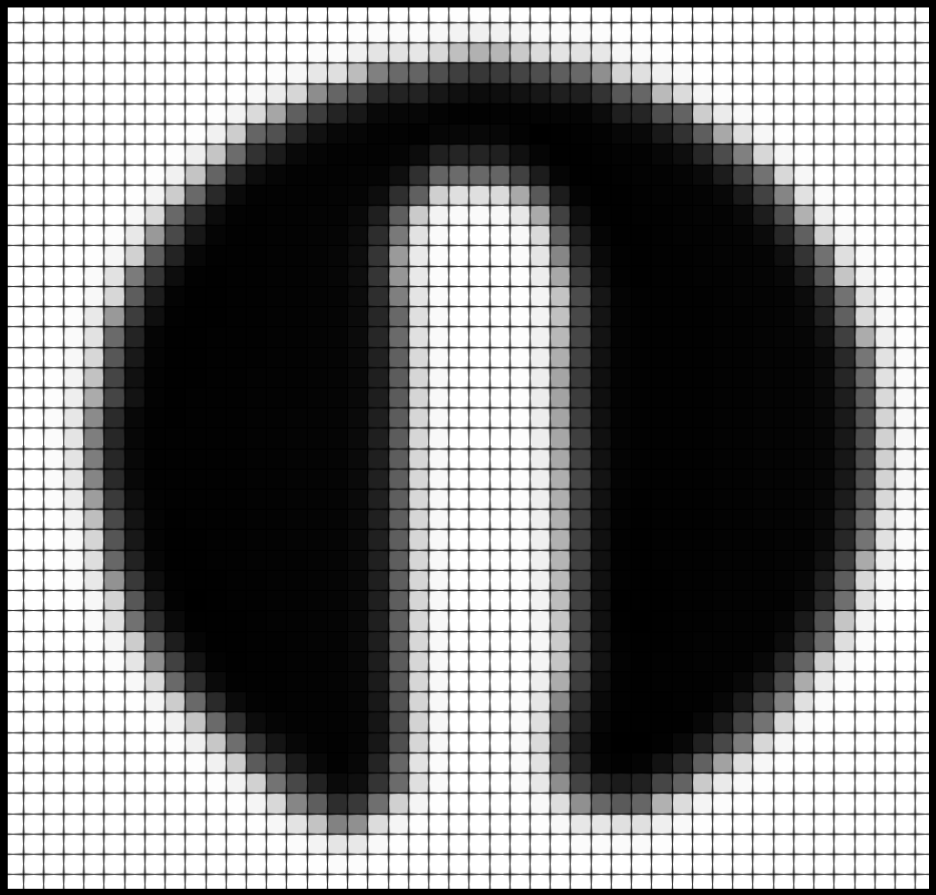}
  \caption{Smoothed field.}
  \label{fig:sub1}
\end{subfigure}
\begin{subfigure}{.43\textwidth}
  \centering
  \includegraphics[width=1.0\linewidth]{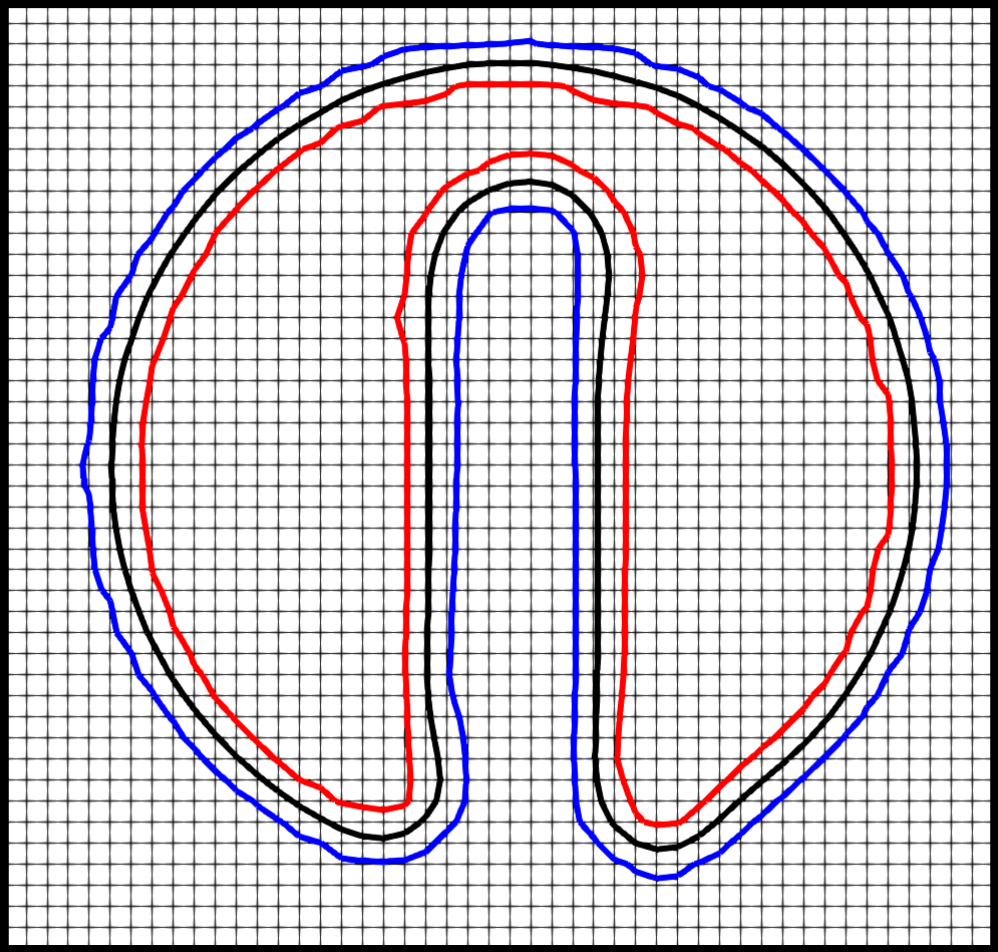}
  \caption{Contours.}
  \label{fig:sub1}
\end{subfigure}
\caption{Numerical results of Zalesak’s slotted disc test for a 100 $\times$ 100 grid after one revolution. The interface is shown for the 0.5 contour (black colour) together with the 0.05 (blue colour) and 0.95 (red colour) contours.}
\label{fig:2DZalesak_contours}
\end{figure}

This case tests the capability of the proposed method to transport a notched disc by a rotating vortex.
The test has been proposed by Zalesak ~\citep{Zalesak1979} and is widely used in the literature for testing interface capturing schemes. A slotted circle centred at (0.5, 0.75) with radius R=0.15 is rotated in a domain $[0,1]^2$. The slot is defined by $|x-0.5| \leq 0.025$ and $y \leq 0.85$. The configuration for the test in Fig.\ref{fig:zalesak_domain} had W = 0.05 and L = 0.25. The velocity field is given by
\begin{linenomath*} 
\begin{align}
\begin{split}
u(x,y,t) = 2\pi (0.5 - y) \\
v(x,y,t) = 2\pi (x - 0.5) 
\end{split}
\end{align} 
\end{linenomath*}
Three levels of grid resolution were used $50 \times 50$, $100 \times 100$ and $200 \times 200$.
After one rotation, the slotted circle returns back to its initial state as in Fig.\ref{fig:zalesak_one_rotation}. The Courant number was 0.3. Fig\ref{fig:2DZalesak_contours} shows the results of a rotated slotted disc after one revolution. 
The most challenging regions of the interface to capture are located at the corners of the slot.
The slot of the circle is distorted during the rotation in those areas, with the error in shape preservation being higher in those regions. The fractional error $E_{r}$ is employed here for consistent comparisons, defined as
\begin{equation}
E_{r}=\frac{\sum_{i}^{ }|\alpha_i-\alpha_{i,exact}|}{\sum_{i}^{}\alpha_{i,exact}}
\end{equation}
The values of $E_{r}$ calculated using different meshes are shown in Table~\ref{table:2DSlottedDisk_Ea}. 
The results were comparable to other VOF geometrical approaches using the THINC/LS scheme with high-order surface polynomials and were generally more accurate than the second-order THINC-LS (P2) and the forth-order THINC-LS (P4) schemes with more than 50$\%$ lower error. The difference with the solver interFoam with the algebraic method in ~\citep{Deshpande2012} is higher and the error of our method here has an order of magnitude less error. 
 
\begin{table}
\caption{$E_{r}$ error of Zalesak rotation test after one rotation and comparison with other methods.}
\centering
\label{table:2DSlottedDisk_Ea}
\begin{tabular}{l l l l}
\hline
Authors & $50^{2}$ & $100^{2}$ & $200^{2}$ \\
\hline

THINC-LS (P1) ~\citep{Qian2018} & 1.23$\times 10^{-1}$ & 3.37$\times 10^{-2}$ & 1.19$\times 10^{-2}$  \\

THINC-LS (P2) ~\citep{Qian2018} & 6.40$\times 10^{-2}$ & 1.47$\times 10^{-2}$ & 7.04$\times 10^{-3}$  \\

THINC-LS (P4) ~\citep{Qian2018} & 3.64$\times 10^{-2}$ & 9.98$\times 10^{-3}$ & 3.99$\times 10^{-3}$  \\

interFoam ~\citep{Deshpande2012} & - & 9.22$\times 10^{-2}$ & 3.92$\times 10^{-2}$  \\
                   
Presented method & 1.38$\times 10^{-2}$ & 5.21$\times 10^{-3}$ & 1.64$\times 10^{-3}$  \\

\hline
\end{tabular}
\end{table}

\subsection{Two-dimensional rising bubble}
\begin{figure}[h]
    \vspace{6pt}
    \centering
    \includegraphics[scale=0.24]{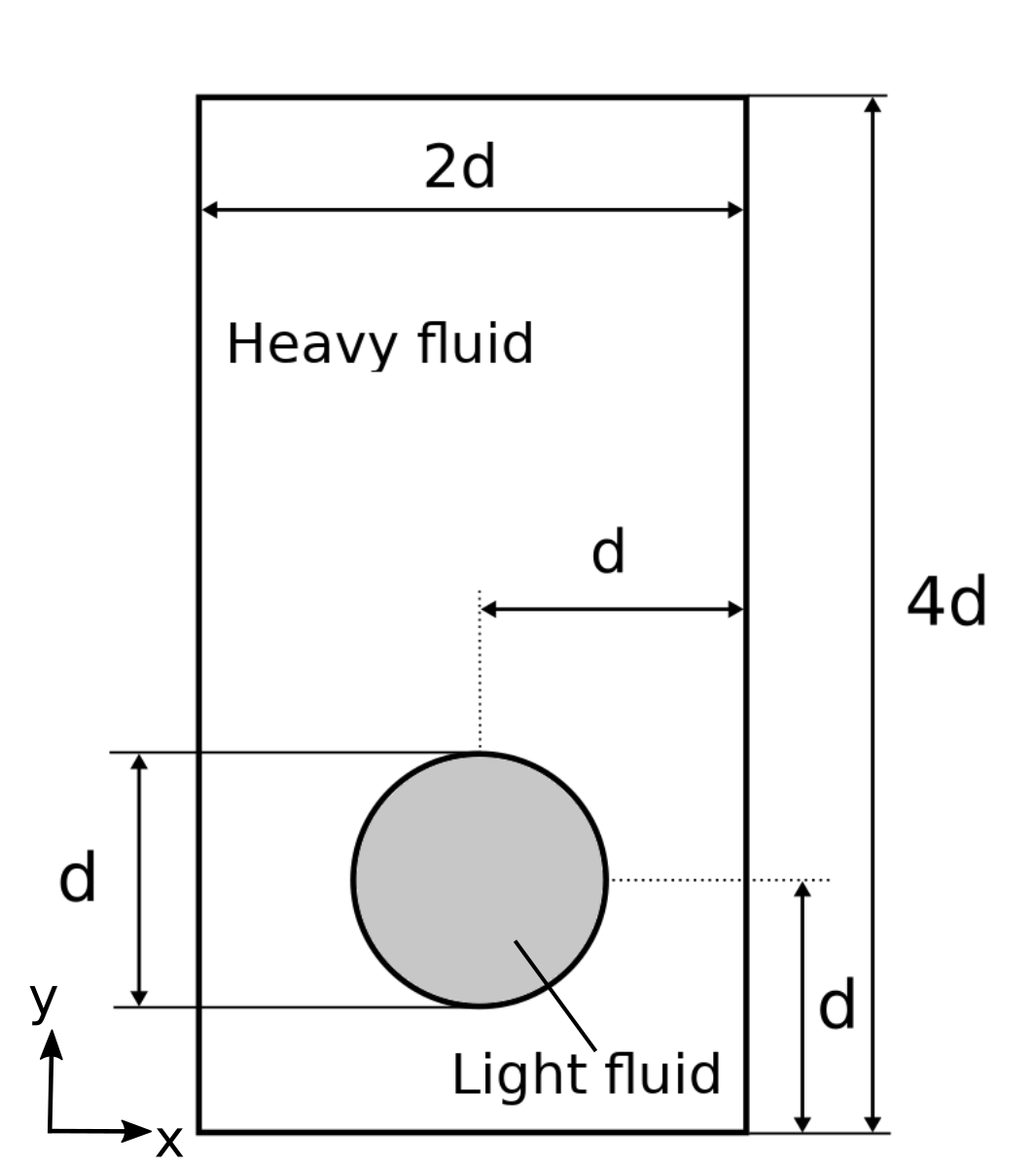}
       \centering
    \caption{Domain for the two-dimensional rising bubble test.}
    \label{fig:rising_bubble_set_up}
\end{figure}

\begin{figure}[H]
 \vspace{6pt}
\centering
\begin{subfigure}{.49\textwidth}
  \centering
  \includegraphics[width=1.05\linewidth]{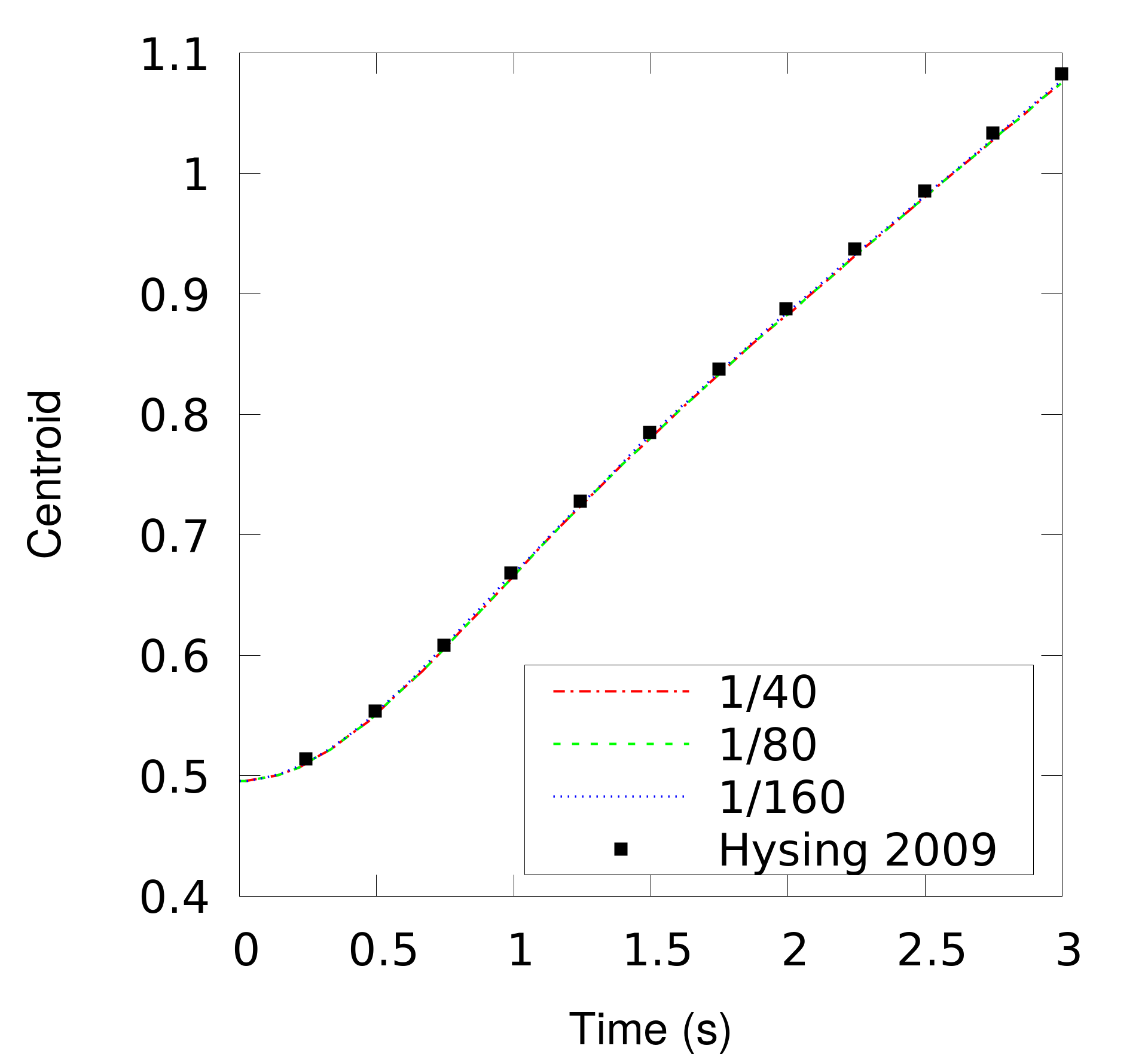}
  \caption{Centroid.}
  \label{fig:sub1}
\end{subfigure}
\begin{subfigure}{.49\textwidth}
  \centering
  \includegraphics[width=1.05\linewidth]{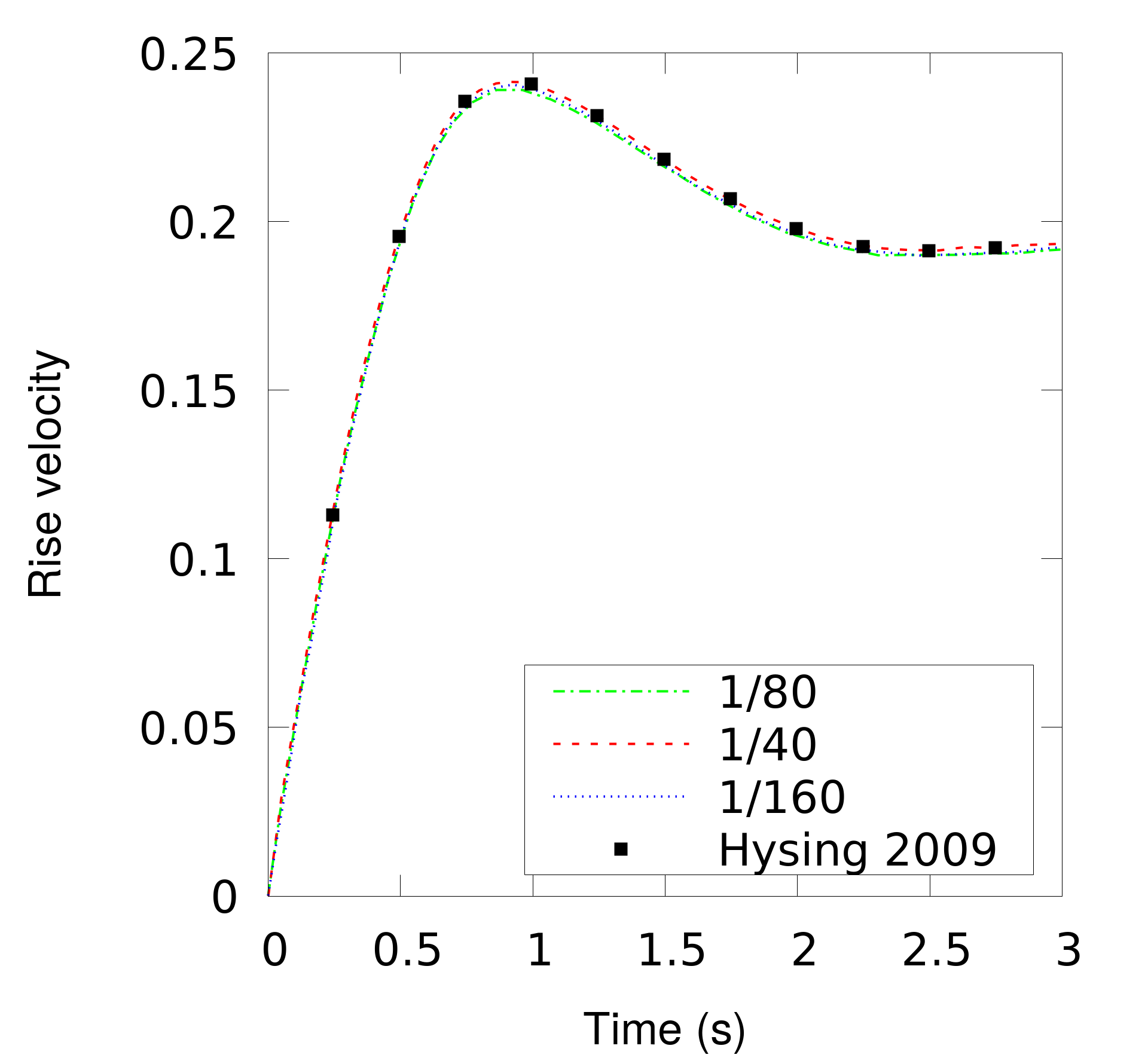}
  \caption{Rise velocity.}
  \label{fig:sub1}
\end{subfigure}
\begin{subfigure}{.49\textwidth}
  \centering
  \includegraphics[width=1.05\linewidth]{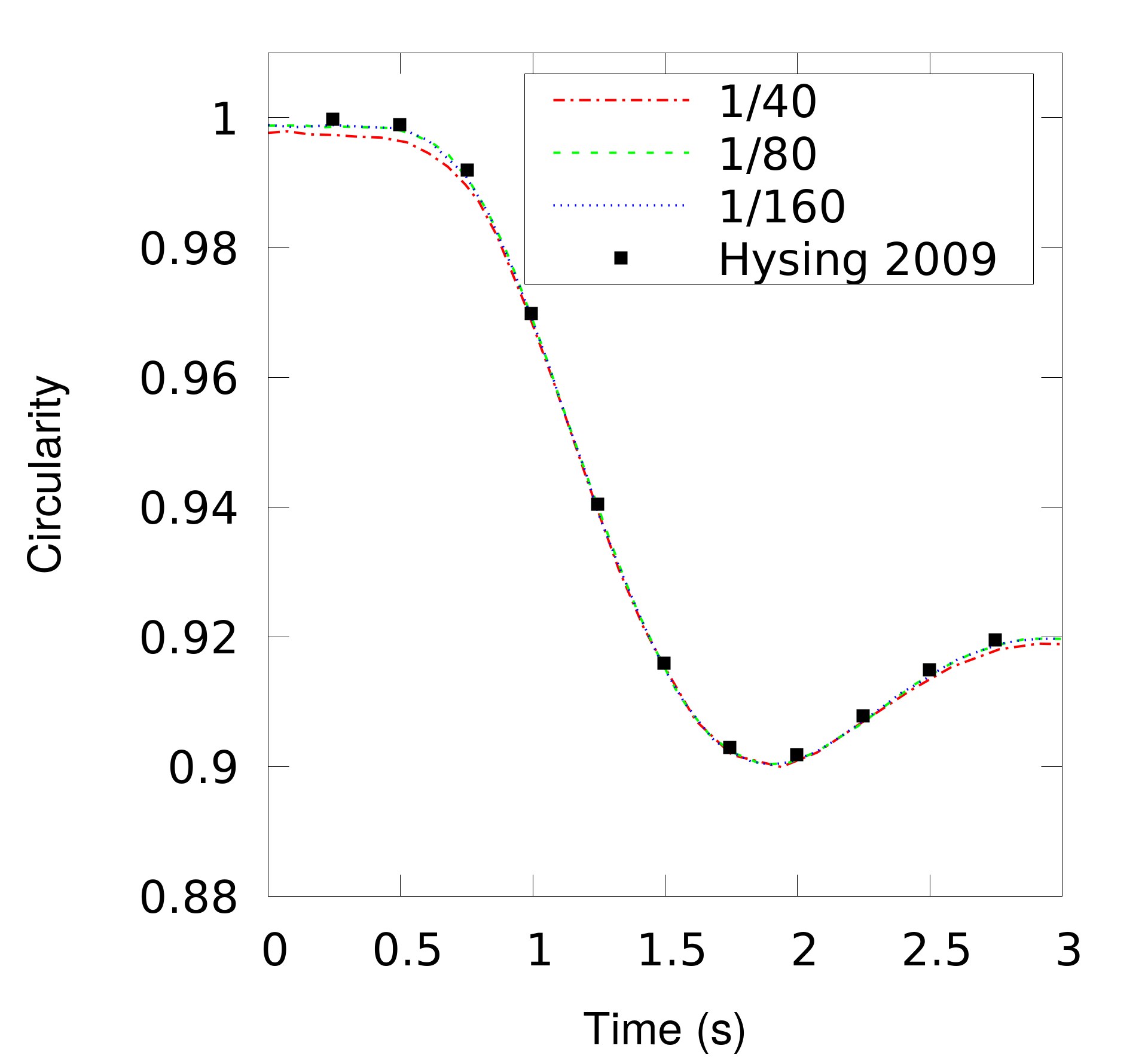}
  \caption{Circularity.}
  \label{fig:sub2}
  \end{subfigure}
\begin{subfigure}{.49\textwidth}
  \centering
  \includegraphics[width=0.6\linewidth]{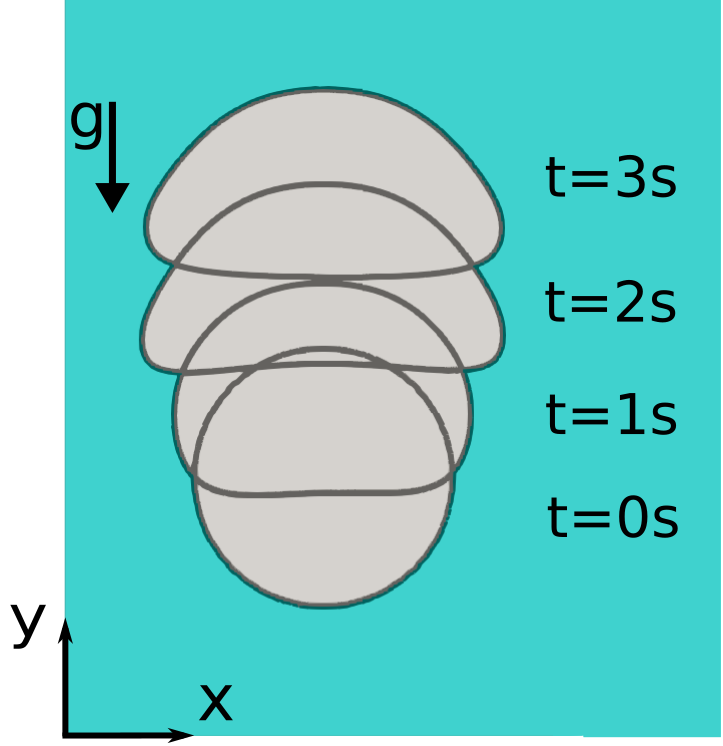}
  \caption{Bubble shape evolution.}
  \label{fig:sub2}
  \end{subfigure}
\caption{Evolution of rising bubble benchmark quantities through time. The results are compared with the results in ~\citep{Hysing2009}.}
\label{fig:2DrisingBubble_yc_vc_zeta}
\end{figure}

In this test a gaseous bubble with volume $\Omega_{1}$ is placed in a column filled with a heavier fluid of volume $\Omega_{2}$ and rises under the effect of gravity.
The bubble rises gradually at the top of the domain and deforms changing shape. 
The test has been proposed by ~\citep{Hysing2009} who presented quantitative benchmark quantities in terms of the mass-centre of the bubble, the rise velocity and its circularity.
Initially at t=0s, the bubble is placed at (0.5, 0.5) in a rectangular domain $[2d, 4d]$ as shown in Fig.\ref{fig:rising_bubble_set_up} and has diameter d = 0.5 units.
No-slip boundary conditions were applied at the top and the bottom of the column and a free-stream boundary condition at the vertical walls.
The velocity is initially zero in the domain and inside the bubble the pressure is set as constant from hydrostatic theory. 
The physical properties for the fluids are listed in Table~\ref{table:2DRisingBubble_phys_properties}.
Different grid resolutions were used for the study: $\Delta x /40, \Delta x /80, \Delta x /160$. 
The Reynolds ($Re= \rho_2 U_T d/\mu_2$) and E\"otv\"os numbers ($Eo = gd^2 \Delta \rho_2 / \sigma$) were Re=35 and Eo=10 for this study. Here, $U_T=\sqrt{gd}$ is the gravitational velocity.
The centre of mass ($y_{c}$), the rise velocity of the bubble ($v_{c}$) and its circularity (sphericity in 3D) ($\zeta$) were defined by \citep{Hysing2009} as
\begin{linenomath*}
\begin{align}
\begin{split}
y_{c}= \frac{\int_{\Omega_1}ydV}{\int_{\Omega_1}dV} \\
v_{c}= \frac{\int_{\Omega_1}vdV}{\int_{\Omega_1}dV} \\
\zeta = \frac{\pi d}{\Pi}
\end{split}
\end{align}  
\end{linenomath*}
where $\Pi$ is the perimeter of the bubble, $y$ the coordinate value and $v$ is the vertical velocity. 
The benchmark quantities are shown in Fig.\ref{fig:2DrisingBubble_yc_vc_zeta} for the different grid resolutions. 
There was good agreement with the results in \citep{Hysing2009} and \citep{Hysing2012} for $x_{c}$, $v_{c}$ and $\zeta$ for the different grid resolutions. 
The bubble loses its initial circular shape for which $\zeta$ =1 at t=0s, and under the forces exerted on the bubble starts to deform. 
After approximately t=1.95s the bubble becomes more flattened and obtains its minimum circularity (approximately $\zeta$ =0.9 at t=2s). At that point and while the centre mass coordinate value increases linearly with time, the circularity starts to increase, under the influence of surface tension force $F_{\sigma}$ which prevents the bubble disintegrating \citep{Balcazar2016}. At the same time for $t \in [2s,3s]$ the rise velocity remained relatively constant with $v_c$ =0.2m/s. The algorithm calculates the surface tension force with the new $\psi$ after it is corrected to satisfy the volume fraction from VOF. Since a segregated approach is used for solving Navier-Stokes equations, a pressure-velocity algorithm is used, which is the Pressure Implicit with Splitting of Operator (PISO) by \citep{Issa1986}. 
When using the PISO algorithm, a pressure equation is solved according to the number specified by a predefined number of times. Here the number of times the algorithm solved the pressure equation and momentum corrector in each step were set to 2. 

\subsection{Three-dimensional rising bubble}
\begin{figure}[h]
    \vspace{6pt}
    \centering
    \includegraphics[scale=0.15]{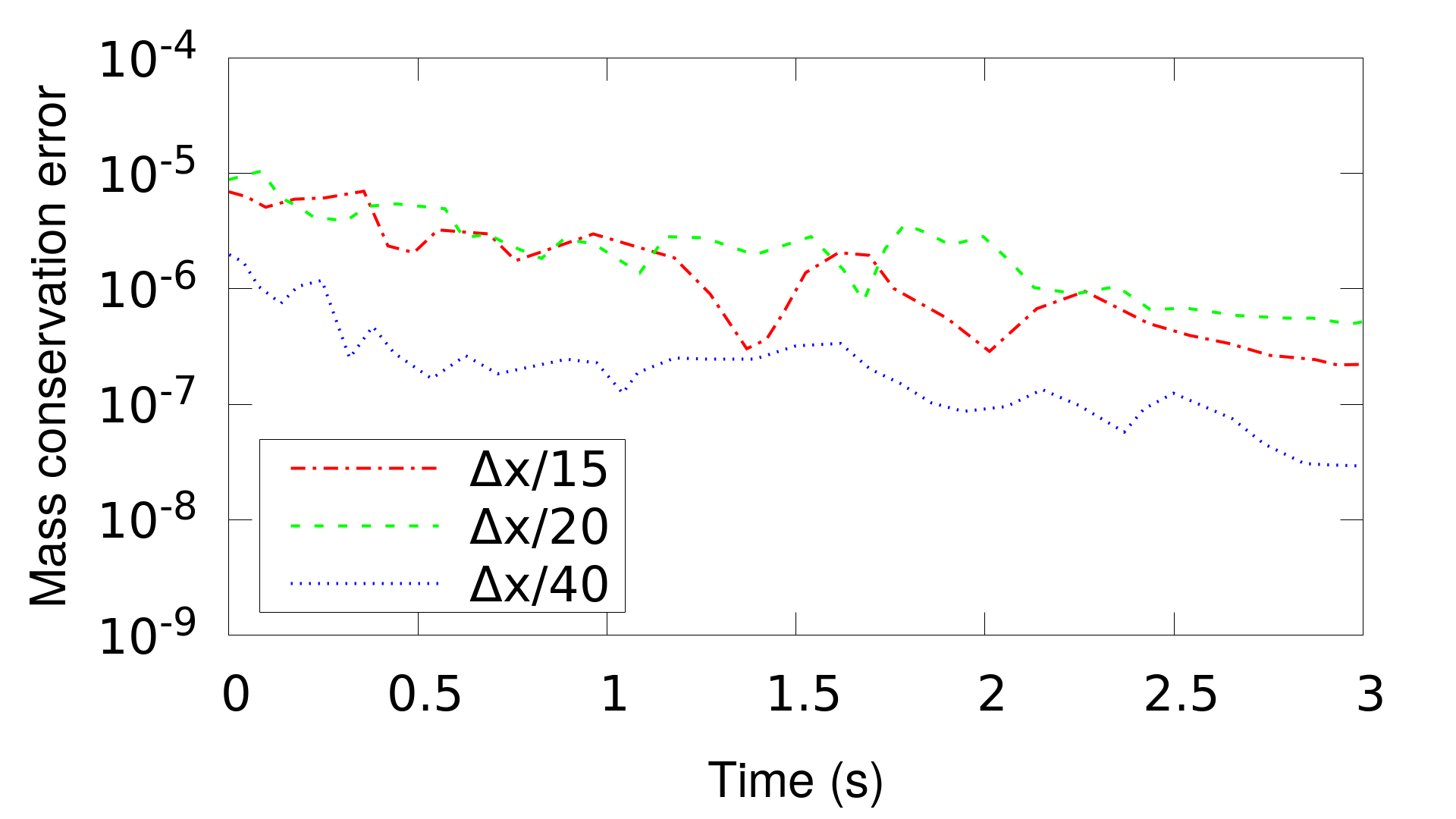}
       \centering
    \caption{Mass conservation error for three-dimensional rising bubble case for the three levels of refinement.}
    \label{fig:3DrisingBubble_mass_conservation}
\end{figure}

A rising bubble is tested next in three-dimensions following the test case in \citep{Balcazar2016}. 
A bubble is placed in a cylindrical domain with diameter $8d$ and height $8d$. 
The distance from the bottom of the domain to the centre of the sphere was $1.5d$.
For the physical parameters of the test, it was $\mu_1/\mu_2 =0.01$ and $\rho_1/\rho_2 =0.01$.
Four meshes with a grid size $\Delta x=d/15, 20, 30, 40$ were used.
The error in Reynolds number and circularity is calculated based on the finest mesh (exact). These are defined as $E_{Re}= (Re - Re_{exact})/Re_{exact}$ and $E_{\zeta}=(\zeta - \zeta_{exact}) / \zeta_{exact}$.
Table~\ref{table:3DRisingBubble} shows the values for $E_{Re}$ and $E_{\zeta}$ compared to the results of the CLSVOF method of \citep{Balcazar2016} and ILSVOF in \citep{Lyras2020} showing better accuracy for the different mesh resolutions. In particular, the error in the other two methods were approximately two times higher than the error obtained with the method here.
The time history of the error in mass conservation considering the mass initially in the domain $\Omega1 \cup \Omega2$, $M(0)$ is calculated as in \citep{Balcazar2016} according to $(M(t)-M(0))/M(0)$, where $M(t)= \int_{\Omega1 \cup \Omega2} \alpha (t) dV$.  
The error is shown in Fig.\ref{fig:3DrisingBubble_mass_conservation} for the various grid resolutions and remained reasonably low during the deformation of the bubble approaching $10^{-6}$ or lower at the end of the simulations. 
  
\begin{table}
\caption{Parameters considered for the rising gas bubble problem.}
\centering
\label{table:2DRisingBubble_phys_properties}
\begin{tabular}{l l l l l l}
\hline
$\rho_1$ & $\rho_2$ & $\mu_1$ & $\mu_2$ & g & $\sigma$ \\
\hline

100$kg / m^{3}$ & 1000$kg / m^{3}$ & 1$kg / m \cdot s $ & 10$kg / m \cdot s $ & 0.981$m/s^2 $ & 24.5 $N/m$  \\

\hline
\end{tabular}
\end{table}

\begin{table}
\caption{Error for Re and sphericity of the bubble in the three-dimensional rising bubble case. Three different structured meshes are used and the results are compared with the CLSVOF method of ~\citep{Balcazar2016}.}
\centering
\label{table:3DRisingBubble}
\begin{tabular}{l c c c c c c c c}
\toprule
\multirow{3}{*}{Mesh size} & \multicolumn{2}{c}{Presented method} & \multicolumn{2}{c}{CLSVOF method of ~\citep{Balcazar2016}} & \multicolumn{2}{c}{ILSVOF method ~\citep{Lyras2020}}\\ 
\cmidrule{2-8}
  & $E_{Re}$ & $E_{\zeta}$ & $E_{Re}$ & $E_{\zeta}$ & $E_{Re}$ & $E_{\zeta}$ &\\ 
\midrule
$\Delta x /15$ & 0.00175 & 0.0053 & 0.00341 & 0.0118 & 0.00363 & 0.0126 &\\

$\Delta x /20$ & 0.00149 & 0.0032 & 0.00339 & 0.0074 & 0.00314 & 0.0063 &\\

$\Delta x /30$ & 0.000083 & 0.0019 & 0.00017 & 0.0042 & - & - &\\

\bottomrule
\end{tabular}
\end{table}

\section{Conclusions}
A conservative level set approach with a correction step using volume of fluid method is presented for simulating two-phase flows in the finite volume framework. The interface capturing is utilised using a resharpening method that conserves mass avoiding smearing of the interface.
The mass fluxes are calculated based on the new smoothed volume fraction that is corrected.
The proposed method supports different types of grids and allows for various options for choosing the VOF advection.  
We show here that with this mass-conservation method for advecting level-set using a simple re-initialisation and the flux update algorithm, one might generally be able to avoid using dedicated re-initialisation schemes for re-sharpening level set. 

The numerical results for highly deformed interfaces with moving and changing topology showed reasonable mass conservation error and accuracy. The mass loss remained low for long time simulations due to the mass-correction step based on the coupling with the volume of fluid method. 
Based on the numerical tests, the methodology can be used for accurately simulating flows with immiscible fluids with surface tension and flows with strong interface deformations.

More tests for improving the calculation of the level set function during the re-initialisation could be used considering algorithms that guarantee at least second-order accuracy for the re-sharpening of the level set regardless of the mesh type. The method uses some predefined parameters that depend on the thickness of the interface and a sensitivity analysis could provide more information for its impact of the correction of $\psi$. 
    
\section*{Acknowledgments}
This work was supported by Heart Research UK [RG2670/18/21] and the Wellcome/EPSRC Centre for Medical Engineering [WT 203148/Z/16/Z]. 
\bibliographystyle{model6-num-names}
\bibliography{mybibfile}

\end{document}